\RequirePackage[2020-02-02]{latexrelease}
\documentclass[aps,prb]{revtex4}
\usepackage{amsmath,amssymb}
\usepackage{graphicx}
\usepackage{dcolumn}
\usepackage{bm}
\usepackage{color}
\usepackage{relsize}
\newcommand{\mbb}{\mathbb}
\newcommand{\mc}{\mathcal}

\newcommand{\tet}{\texttt}

\begin{document}
\title{Dynamical polarization function, plasmons, their damping and collective effects \\ in semi-Dirac bands}
\author{
Gabrielle Ross-Harvey$^{1}$,
Andrii Iurov$^{1}$\footnote{E-mail contact: aiurov@mec.cuny.edu, theorist.physics@gmail.com
},
Liubov Zhemchuzhna$^{1}$,
Godfrey Gumbs$^{2,3}$, 
and
Danhong Huang$^{4}$
}

\affiliation{
$^{1}$Department of Physics and Computer Science, Medgar Evers College of City University of New York, Brooklyn, NY 11225, USA\\ 
$^{2}$Department of Physics and Astronomy, Hunter College of the City University of New York, 695 Park Avenue, New York, New York 10065, USA\\ 
$^{3}$Donostia International Physics Center (DIPC), P de Manuel Lardizabal, 4, 20018 San Sebastian, Basque Country, Spain\\ 
$^{4}$Space Vehicles Directorate, US Air Force Research Laboratory, Kirtland Air Force Base, New Mexico 87117, USA\\ 
}

\date{\today}

\begin{abstract}
We have calculated the dynamical polarization, plasmons and damping rates in semi-Dirac bands (SDB’s) with zero band gap and half-linear, half-parabolic low-energy spectrum. The obtained plasmon dispersions are strongly anisotropic and demonstrate some crucial features of both two-dimensional electron gas and graphene. Such gapless energy dispersions lead to a localized area of undamped and low-damped plasmons in a limited range of the frequencies and wave vectors. The calculated plasmon branches demonstrate an increase of their energies for a finite tilting of the band structure and a fixed Fermi level which could be used as a signature of a specific tilted spectrum in a semi-Dirac band.  
\end{abstract}

\maketitle

\section{Introduction} 
\label{sec1}
Since the discovery of graphene and "graphene miracle", all the two-dimensional materials with a Dirac cone and investigating their various electronic properteis 
have become a crucial part of condensed matter physics. These materials include recently discovered $\alpha-\mc{T}_3$ model with a flat band, \,\cite{tabert2013valley, iurov2020klein, islam2023properties, iurov2019peculiar,tamang2023probing,islam2023effect,gorbar2019electron,illes2017klein} anisotropic and tilted 1T'-MoS$_2$, \cite{yan2023highly, iurov2020quantum, tan2021anisotropic, gomes2021tilted} semi-Dirac materials \cite{carbotte2019optical,islam2018driven,xiong2023optical,mondal2022topology} and materials with Rashba spin-orbit coupling. \,\cite{shitrit2013spin, shih2022blocked, wang2005plasmon} An anisotropy and energy 
gap in the band structure of Dirac cone materials could be also induced by applying external off-resonance irradiaton. \,\cite{kristinsson2016control,kibis2010metal}

\medskip 

Plasmons, or collective quantum density oscillations in an interacting
electron system represent one of the most important directions in low-dimensional physics and have been investigated in great depth for 
graphene\,\cite{politano2014plasmon,hwang2007dielectric,polini2008plasmons,wunsch2006dynamical}, graphene with a finite bandgap and buckled honeycomb lattices
\,\cite{pyatkovskiy2008dynamical,tabert2014dynamical}, graphene-based heterostructures \,\cite{iurov2017effects,yao2018broadband,woessner2015highly,li2017first,iurov2017controlling} at both zero and finite temperatures, \,\cite{sarma2013intrinsic,iurov2017thermal} double and multi-layer systems \,\cite{sarma1981collective,iurov2020many} as well as in specific low-dimensional structures, such as fullerenes \,\cite{henrard1999electron, gumbs2014strongly, solov2005plasmon, ju1993excitation} and nanoribbons. \,\cite{brey2007elementary,karimi2017plasmons, gomez2016plasmon, iurov2021tailoring, fei2015edge} Specifically, there has been 
a large number of papers intended to study the plasmons and electronic transport in the presence of a magnetic field.\,\cite{yan2012infrared, 
roldan2009collective, roldan2011theory, balassis2020magnetoplasmons, dutta2023intrinsic, tamang2023orbital, nimyi2022landau, oriekhov2023size} 

\par 
A considerable attention has been also directed to how the plasmons are excited, \,\cite{farhat2013exciting, brongersma2015plasmon, vinogradov2018exciting}
as well as they lifetime and instability. \,\cite{simon1983inhomogeneous, gumbs2015tunable, petrov2017amplified, koseki2016giant}

\medskip 
It is also important to investigate how the plasmons in any new materials are affected by the most specific and distingushed features of their electronic 
band structure, such as a flat dispersionless band in $\alpha-\mc{T}_3$ materials \,\cite{malcolm2016frequency,tabert2013valley,oriekho2023quantum,oriekhov2020rkky,islam2023role} and plasmons in twisted graphene bilayers. \,\cite{stauber2013optical}

\par
Specifically, the plasmons have been investigated in a large number of newly discovered Dirac and semi-Dirac materials with anisotropy \,\cite{hayn2021plasmons} and tilting (and, possibly, over-tilting),\,\cite{yan2022anomalous, kajita2014molecular} such as screening in 8-Pmmn borophene, \,\cite{sadhukhan2017anisotropic} tilted 1T'MoS$_2$, \,\cite{balassis2022polarizability} hyperbolic plasmons in massive tilted two-dimensional Dirac materials with linear dispersions in which the mass is induced by a bandgap  \,\cite{mojarro2022hyperbolic} optical properties in tilted Dirac systems, \,\cite{mojarro2021optical} kinks in the plasmons in tilted two-dimensional Dirac systems, \cite{jalali2018tilt} hyperbolic plasmon modes in borophene, \,\cite{torbatian2021hyperbolic, sadhukhan2020novel} as well as in  triple component fermionic systems. \,\cite{dey2022dynamical}


\medskip 
\medskip 
The remaining part of the present paper is organized as follows. In section \ref{sec2}, we analyze the low-energy Hamiltonian of semi-Dirac bands and the resulting energy dispersions, as well as derive the corresponding wave functions. We discuss the peculiar properties of the energy spectrum in SDB’s -- half-linear half-parabolic in all detail, and find the doping density required to achieve a certain Fermi level.  Next, in Section \ref{sec3}, we consider the polarization function for semi-Dirac bands, specific overlap factors, dielectric function and the plasmon dispersions together with their damping rates and provide a detailed discussion of our obtained numerical results. Finally, the concluding remarks are made in section \ref{sec4}.

\medskip
\par
\begin{figure} 
\centering
\includegraphics[width=0.55\textwidth]{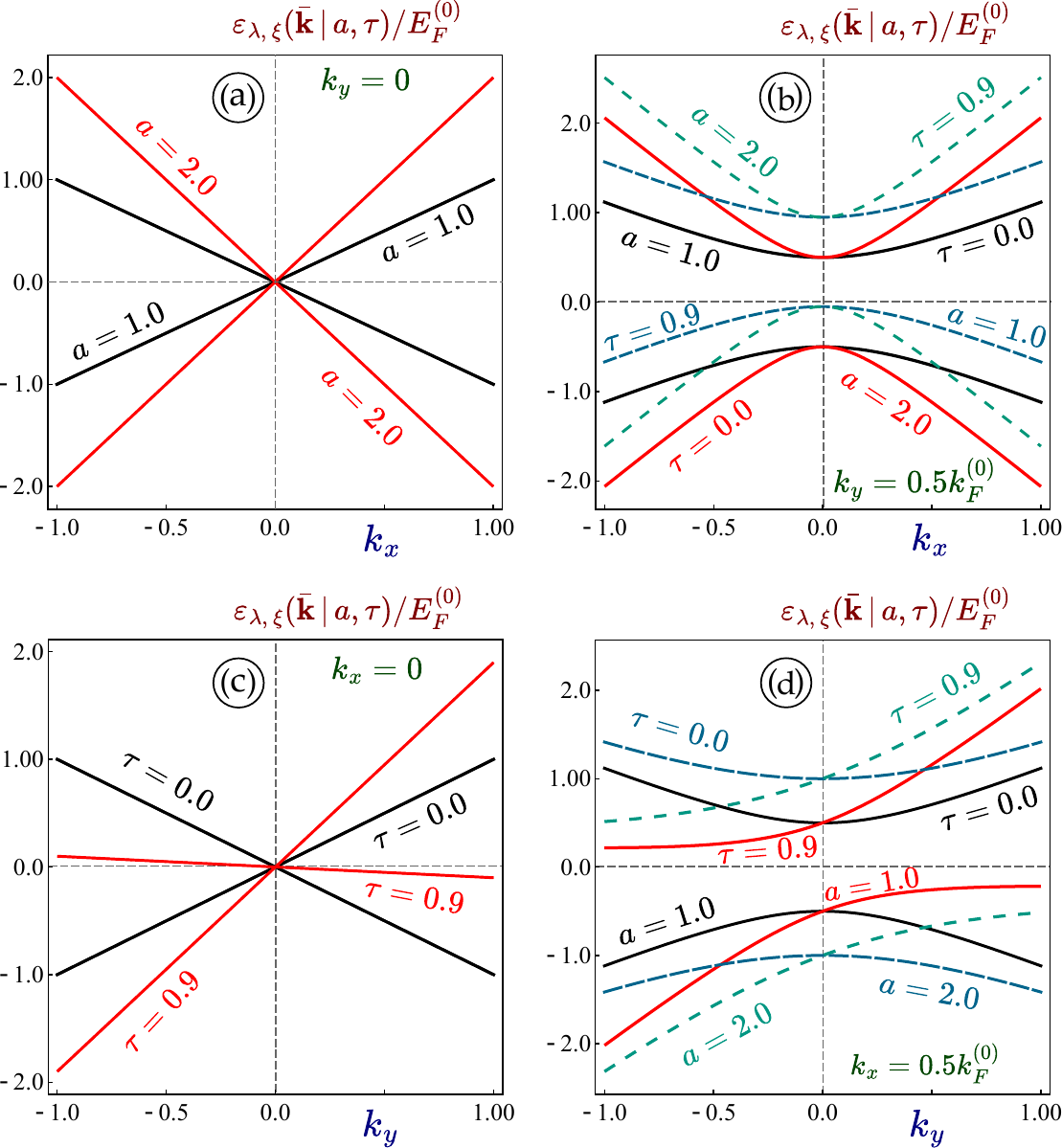}
\caption{(Color online) Energy dispersions  $\varepsilon_{\lambda, \xi} (\bar{\bf k} \, \vert \, a, \tau)$ in semi-Dirac bands plotted as functions of 
$k_x$ (upper panels $(a)$ and $(b)$) and $k_y$ (lower panels $(a)$ and $(b)$). The two left panels $(a)$ and $(c)$ are plotted for the zero transverse 
momentum component, and plots $(b)$ and $(d)$ - for its finite value $0.5 \, k_F^{(0)}$. Each curve corresponds to the various values of parameters $\tau$ and 
$a$ of energy dispersions \eqref{disp01}, as labeled.    
}
\label{FIG:1}
\end{figure}

\medskip
\par
\begin{figure} 
\centering
\includegraphics[width=0.8\textwidth]{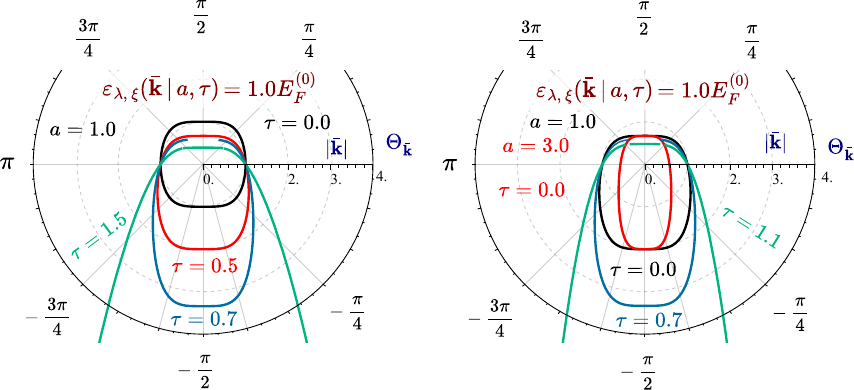}
\caption{(Color online) Energy dispersions $\varepsilon_{\lambda, \xi} (\bar{\bf k} \, \vert \, a, \tau)$ in semi-Dirac bands. Horizontal constant-energy 
cuts of the three-dimensional dispersion plots of $\varepsilon_{\lambda, \xi} (\bar{\bf k} \, \vert \, a, \tau)$ demonstrate the Fermi surface -- an interface 
between the occupied and free states at zero temperature -- for various values of tilting parameter $\tau$. Panel $(a)$ is a polar plot for the $E_F = \text{const}$-cut of semi-Dirac bands with different $\tau$, while plot $(b)$ demonstrates the effect of both $\tau$ and $a$. 
}
\label{FIG:2}
\end{figure}

\medskip
\par
\begin{figure} 
\centering
\includegraphics[width=0.95\textwidth]{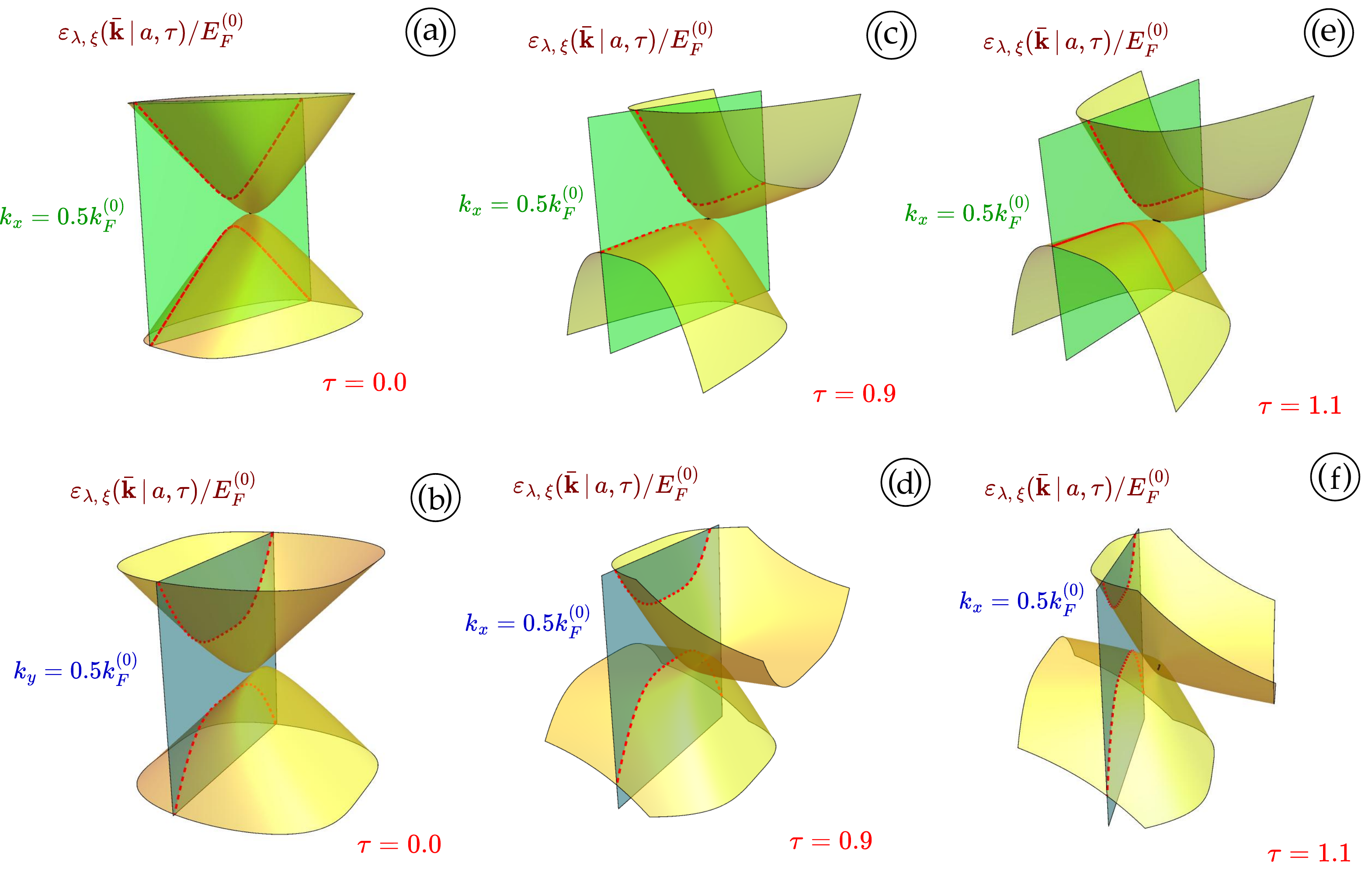}
\caption{(Color online) 
Energy dispersions $\varepsilon_{\lambda, \xi} (\bar{\bf k} \, \vert \, a, \tau)$ in semi-Dirac bands. Vertical $k_x = \text{const}$ and $k_y  = \text{const}$ cuts of the three-dimensional dispersion plots are used to demonstrate the anisotropy, tilting and the specific spectra of the energy band structure in the
$k_x$ and $k_y$ directions. 
}
\label{FIG:3}
\end{figure}

\medskip
\par
\begin{figure} 
\centering
\includegraphics[width=0.95\textwidth]{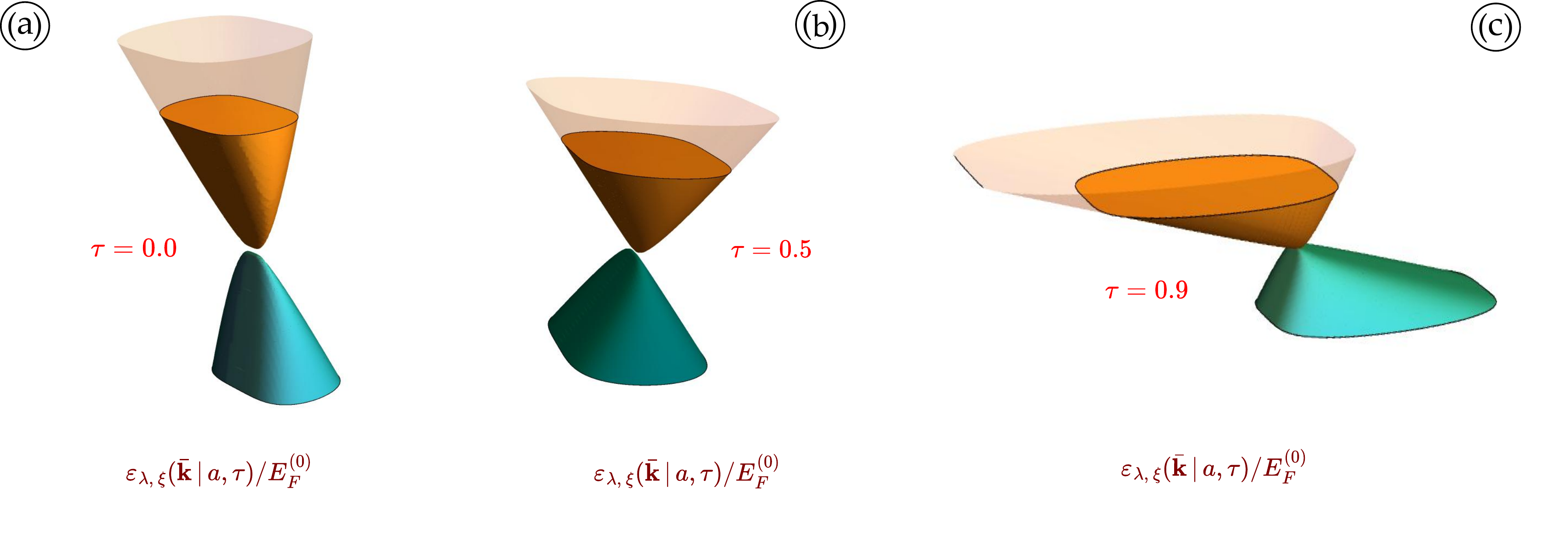}
\caption{(Color online) 
Fermi level and the Fermi surface for zero ($\tau = 0$), finite ($0 < \tau < 1$) and critical $(\tau \rightarrow 1)$ tilting. Three-dimensional plots of the 
energy dispersions $\varepsilon_{\lambda, \xi} (\bar{\bf k} \, \vert \, a, \tau)$ for a finite Fermi level $E_F^{(0)}$ reveal increased an Fermi surface for 
a finite tilting $\tau$, and an infinite one for over-critical tiling $\tau > 1$. 
}
\label{FIG:4}
\end{figure}

\section{General formalism}
\label{sec2}

The low-energy dispersions of semi-Dirac bands (SDB’s) next to the zero-energy Dirac point  are linear in one direction $\backsim v_F k_y$ and quadratic in the other $\backsim (a k_x^2)^2$.  Apart from that, a finite tilting of the energy bands in the $y-$ direction could be also present.

\par 
As a result, for the low-energy states in SDB's we obtain the following Hamiltonian 

\begin{equation}
\label{hdgg06}
\hat{\mc{H}}_\xi ({\bf k}) = \hbar t \, v_F \, k_y  \hat{\Sigma}_0^{(2)} + a_0 \hbar^2 \, k_x^2  \, \hat{\Sigma}_x^{(2)} + 
\hbar \tau \, v_F \, k_y  \hat{\Sigma}_y^{(2)} \, , 
\end{equation}
where $\hat{\Sigma}_0^{(2)}$ is a $2 \times 2$ unit matrix, $\hat{\Sigma}_x^{(2)}$ and $\hat{\Sigma}_2^{(2)}$ are Pauli matrices, parameter $a_0 = 1/(2 m^\star)$ plays a role of the inverse effective mass.  

\par 
The tilting parameter $\tau$ is essentially the ratio between the Fermi velocities for the diagonal and off-diagonal linear terms in Hamiltonian \eqref{hdgg06} which could be either zero or finite, and even exceed unity; $v_F = 1.0 \cdot 10^6 \, m/s$ is the Fermi velocity in graphene.  

\medskip 
The explicit matrix form of Hamiltonian \eqref{hdgg06} is 

\begin{equation}
\label{ham001}
 \hat{\mc{H}}_\xi ({\bf k} \, \vert \, a, \tau) = \hbar \, \left\{
\begin{array}{c c}
v_F \, \tau \, k_y & \hbar v_F \, a k_x^2 - i  v_F k y \\
\hbar v_F \, a k_x^2 + i  v_F k y & v_F \, \tau \, k_y
\end{array}
\right\} \, ,
\end{equation}
where we used a notation $a = a_0 \hbar$. 

\medskip
The energy spectrum of semi-Dirac bands obtained as the eigenvalues of Hamiltonian \eqref{ham001} in the following form 

\begin{equation}
\label{disp01}
\varepsilon_{\lambda, \xi} (\bar{\bf k} \, \vert \, a, \tau) =  \xi \,\tau \,  v_F k_y
+ \lambda \, \sqrt{(\hbar v_F \, k_y)^2 + (a \, k_x^2)^2} \, . 
\end{equation}

The corresponding wave functions are 

\begin{eqnarray}
\label{wf01}
{\bf \Psi}_{\lambda = \pm 1} (\bar{\bf k} \, \vert \, a, \tau)  & = & \frac{1}{\sqrt{2}} \, \left[
\begin{array}{c}
1 \\
\lambda \, \frac{a \, k_x^2 + i v_F \, k_y}{(v_F \, k_y)^2 + (a \, k_x^2)^2}
\end{array}
\right] \, , 
\end{eqnarray}
meaning that diagonal term $\xi \tau \,  v_F k_y$ has no effect on the wave function which also appears to be valley-degenerate. Introducing a 
vector

\begin{equation}
\label{vecE01}
\bar{\bf \mc{E}} (\bar{\bf k}) = 
\left[
\begin{array}{c}
\mc{E}_x (\bar{\bf k}) \\
\mc{E}_y (\bar{\bf k})
\end{array}
\right]
=
\left[
\begin{array}{c}
\text{Re}(a \, k_x^2 + i v_F \, k_y) \\[0.09cm]
\text{Im}(a \, k_x^2 + i v_F \, k_y) 
\end{array}
\right] 
=
\left(
\begin{array}{c}
a \, k_x^2 \\
v_F \, k_y 
\end{array}
\right) 
\end{equation}
so that

\begin{equation}
\mc{E}  (\bar{\bf k} \, \vert \, a, \tau) = \vert \bar{\bf \mc{E}} \vert  (\bar{\bf k} \, \vert \, a, \tau)
 = \varepsilon_{\lambda, \xi} (\bar{\bf k} \, \vert \, a, \tau) -  \xi \,\tau \,  v_F k_y \, 
\end{equation}
we can introduce an angle $\Theta_{\bar{\bf \mc{E}}} (\bar{\bf k} \, \vert \, a, \tau) = \tan^{-1} \left( \mc{E}_y/ \mc{E}_x \right) = (v_F \, k_y)/(a \, k_x^2)$ and rewrite wave function \eqref{wf01} as 

\begin{eqnarray}
\label{wf02}
{\bf \Psi}_{\lambda = \pm 1} (\bar{\bf k} \, \vert \, a, \tau) & = & \frac{1}{\sqrt{2}} \, \left[
\begin{array}{c}
1 \\
\lambda \, \tet{e}^{i \, \Theta_{\bar{\bf \mc{E}}} (\bar{\bf k} \, \vert \, a, \tau) }
\end{array}
\right] \, .
\end{eqnarray}
We note that a simplified representation \eqref{wf02} of the wave function in semi-Dirac bands is given in terms of an angle $\Theta_{\bar{\bf \mc{E}}} (\bar{\bf k} \, \vert \, a, \tau)$ associated with vector \eqref{vecE01} but not with the components of wave vector $\bar{\bf k}$ directly.

\medskip
\par
\begin{figure} 
\centering
\includegraphics[width=0.6\textwidth]{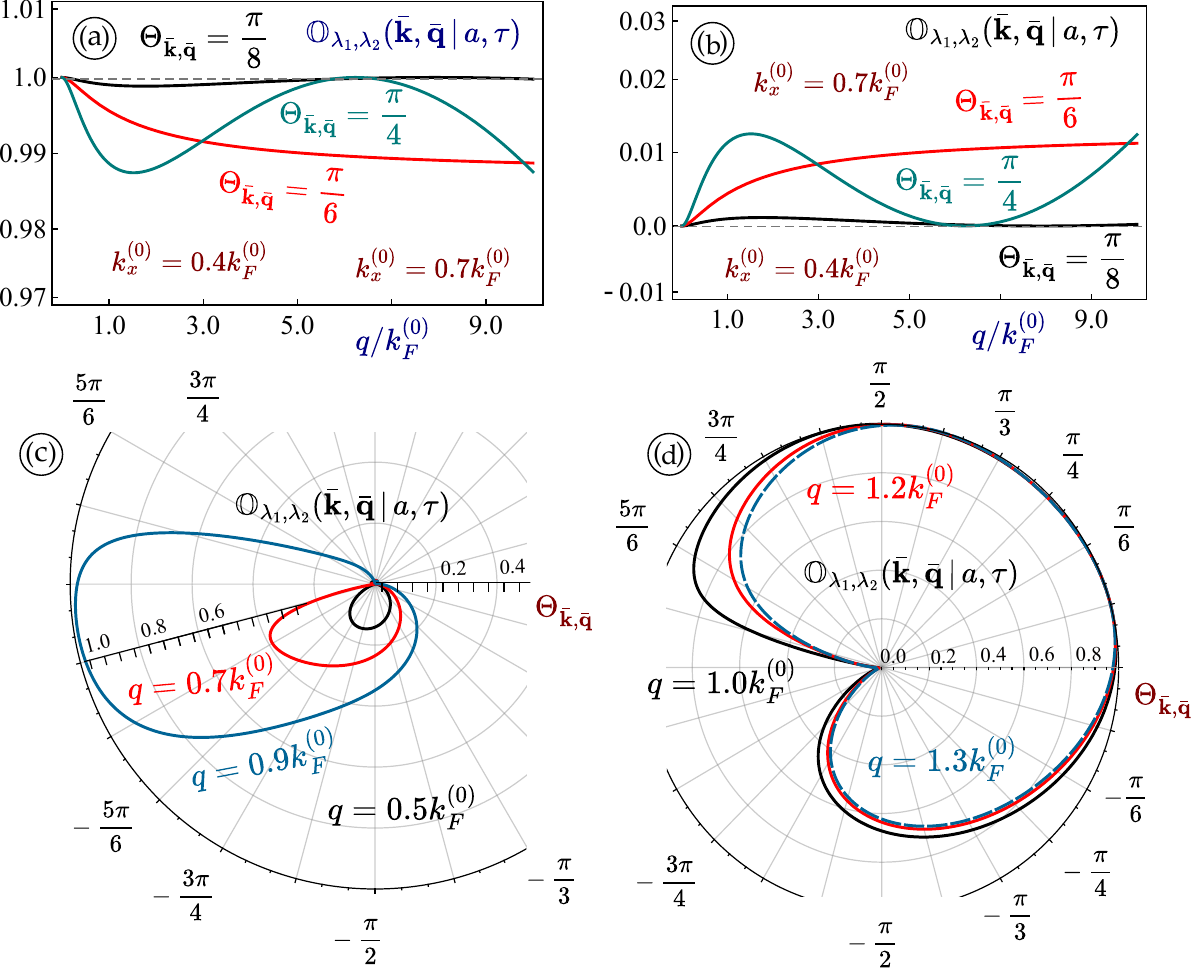}
\caption{(Color online) 
Overlap factors $\mbb{O}_{\lambda_1,\lambda_2} (\bar{\bf k}, \bar{\bf q} \,\vert \, a, \tau)$ for various values of a wave vector $\bar{\bf q}$. All panels
are calculated for a vector $\bar{\bf k}^{(0)} = \left( k_x^{(0)}, k_y^{(0)}\right) = (0.7, 0.4) \, k_F^{(0)}$. Left-hand-side panels $(a)$ and $(c)$ correspond
to {\it intra-band overlaps} with $\lambda_1 \lambda_2 = 1$, and the right ones - to {\it inter-band overlaps} with $\lambda_1 \lambda_2 = - 1$. 
}
\label{FIG:5}
\end{figure}

\medskip
\par
\begin{figure} 
\centering
\includegraphics[width=0.6\textwidth]{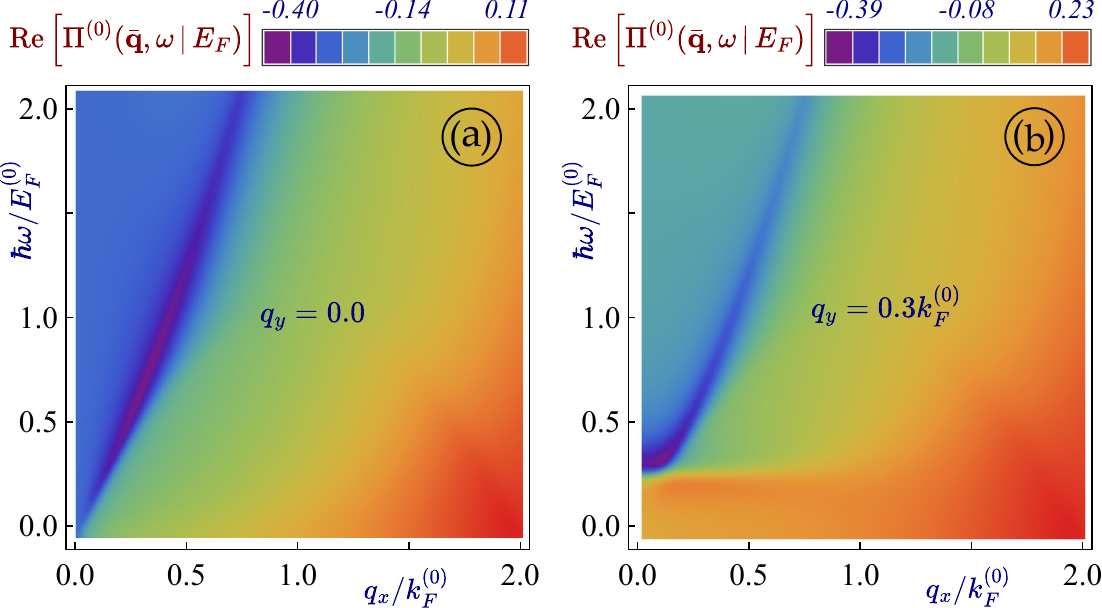}
\caption{(Color online) Real part of the polarization function $\Pi^{(0)}(\bar{\bf q}, \omega\, \vert \, E_F)$ for semi-Dirac bands
as a function of wave vector component $q_x$ and frequency $\omega$. Panel $(a)$ corresponds to $q_y = 0.0$, plot $(b)$ - to $q_y = 0.3 \, k_F^{(0)}$.
}
\label{FIG:6}
\end{figure}

\medskip
\par
\begin{figure} 
\centering
\includegraphics[width=0.6\textwidth]{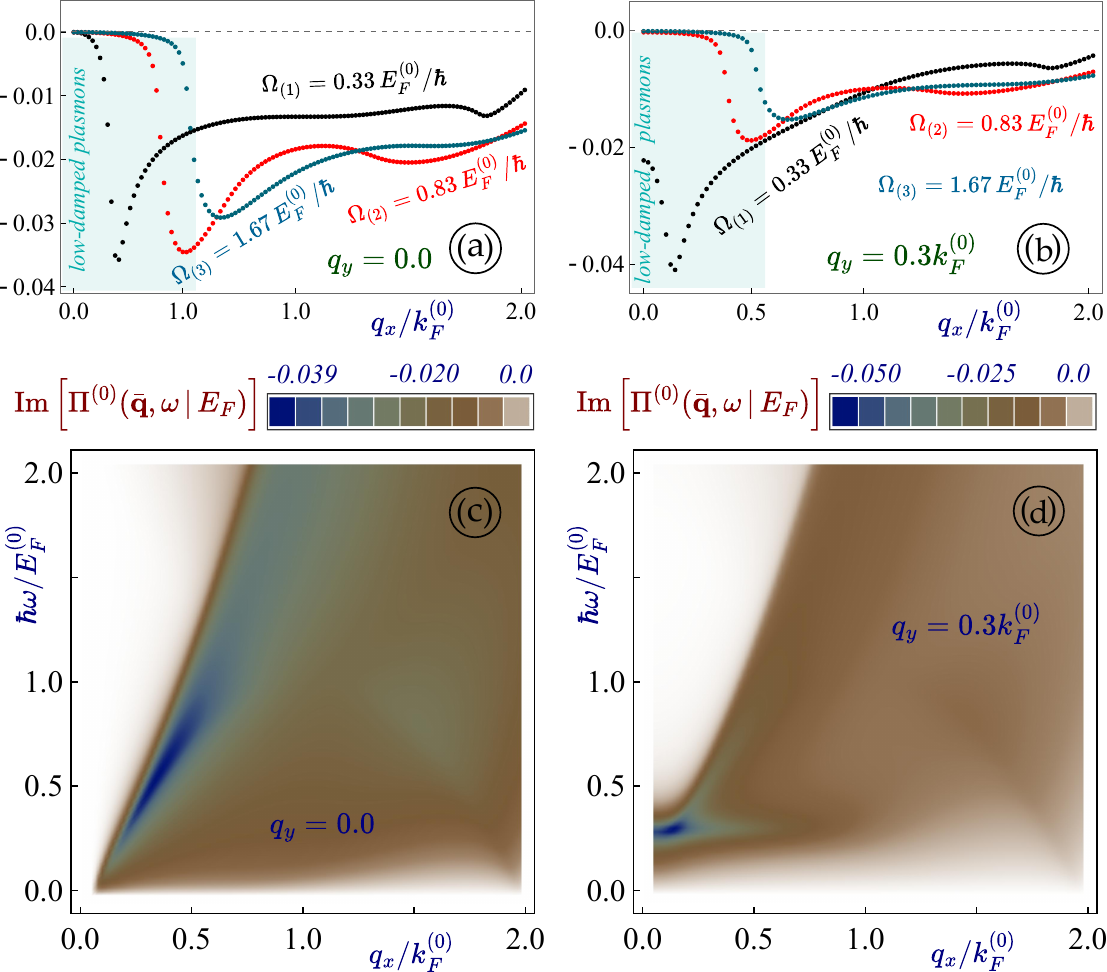}
\caption{(Color online) Particle-hole modes (single-particle excitation regions) obtained the the regions of a finite imaginary part of the 
polarization function $\Pi^{(0)}(\bar{\bf q}, \omega\, \vert \, E_F)$ for semi-Dirac bands
as a function of wave vector component $q_x$ and frequency $\omega$. Panels $(a)$ corresponds to $q_y = 0.0$, plot $(b)$ - to $q_y = 0.3 \, k_F^{(0)}$.
}
\label{FIG:7}
\end{figure}

\medskip
\par
\begin{figure} 
\centering
\includegraphics[width=0.6\textwidth]{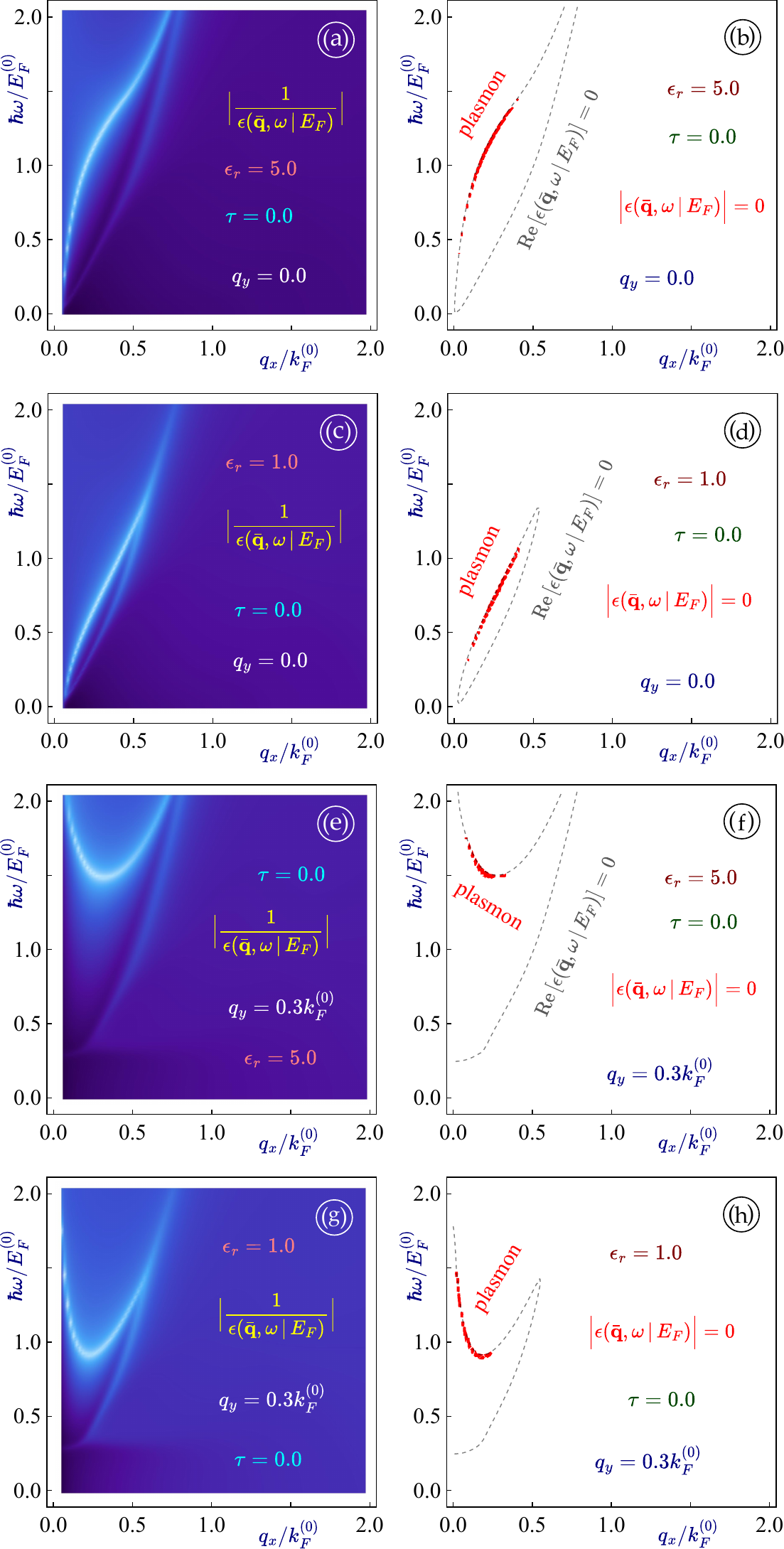}
\caption{(Color online) Plasmon dispersion relation 
for semi-Dirac bands as a function of wave vector component $q_x$ and frequency $\omega$. The left panels $(a)$, $(c)$, $(e)$ and 
$(g)$ represent the density plots of the inverse dispersion function, while the right-hand-side plots $(b)$, $(d)$, $(f)$ and 
$(f)$ - numerically calculated plasmon dispersions obtained as the zeros of the absolute value of the dielectric function \eqref{epsmain}. 
The results presented in various panels correspond to the different values of a component $q_y$ of the wave vector, as labeled.  
}
\label{FIG:8}
\end{figure}

\medskip
\par
\begin{figure} 
\centering
\includegraphics[width=0.6\textwidth]{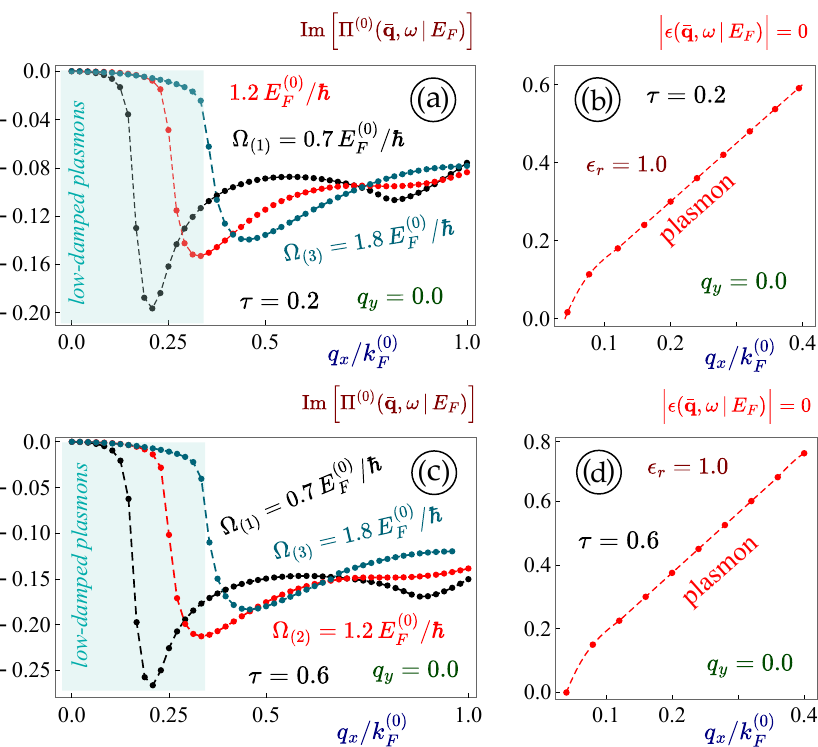}
\caption{(Color online) Plasmons in semi-Dirac bands with a finite tilting $\tau$. The two  left-hand-side panels $(a)$ and $(c)$ demonstrate the imaginary part of the polarization function with a range of wave vectors $q$ with low-damped plasmons. The two right panels $(b)$ and $(d)$ show the corresponding plasmon branches with a zero or a small damping. The two upper plots $(a)$ and $(b)$ are obtained for tilting parameter $\tau = 0.2$, and the two lower ones $(c)$ and $(d)$ -  for $\tau = 0.7$.}
\label{FIG:9}
\end{figure}

\medskip
The most basic and informative two-dimensional plots for the band structure of semi Dirac bands For various  values of  tilting parameters  and inverse effective mass $a_0$ are presented in Fig.~\ref{FIG:1}. As expected to see that if one component of the electron momentum was taken zero, the dependence on the other component is linear. The band structure doesn't reveal any energy band Gap. We also see that a finite value of towel leads over the tilting of spectrum in the $k_y$ direction. 

\medskip
The shape and size of the “horizontal” constant-energy cuts of dispersions  $\varepsilon_{\lambda, \xi} (\bar{\bf k} \, \vert \, a, \tau)$ shown in Fig.~\ref{FIG:2} feature the Fermi surface: a boundary between the occupied and free electronic states of semi-Dirac bands. The size of those surfaces definitely depend on tilting. Once parameter $\tau$ is increased, the surfaces become extended in the $k_y$-direction (corresponding to $\theta_{\bf k } = \frac{\pi}{2}$ and $\frac{3 \pi}{2}$). For $\tau = 1$,  which we are going to refer to as critical tilting, it becomes infinitely large,  as well as for any $\tau > 1$.  The inverse effective mass $a$  makes those Fermi surfaces less circular and more anisotropic. 

\medskip
In Fig.~\ref{FIG:3}, will also plot the the vertical ($k_x = \text{const}$ and $k_y = \text{const}$) cuts of dispersions \eqref{disp01} which reveal all the specific features of the non-trivial band structure in SDB’s,  such as their very specific shapes distinct from those in graphene which  stems from the non-linear dispersions in the $x-$direction. The tilting could be zero,  finite $(0 < \tau < 1)$, critical ($\tau  = 1$) or even over-critical ($\tau  = 1$) making one of the slopes in the $k_y$-direction negative,  as well as substantial anisotropy and overall difference between $k_x$- and $k_y$-dispersions of the seven Dirac bands.

\medskip
We also demonstrate the Fermi surface on Fig.~\ref{FIG:4} showing both occupied and unoccupied states and the interface between them.\footnote{
An informative description with a few recipes for finding an intersection curve for two given surfaces using Wolfram
Mathematica similar to what was used here could be found in https://community.wolfram.com/groups/-/m/t/177994}
It is clearly seen that for the increasing tilting, the surface becomes extended in the $y-$ direction, increases in size and becomes unbounded and infinite for $\tau \geq 1$. For the critical or over-critical tilting $\tau$,  it is possible to observe the Fermi surface in both valence and conduction bands at the same time in contrast to the most known Dirac materials. A finite-size Fermi surface is also possible even for a zero doping $E_F = 0$.

\section{Polarization function and plasmons in semi-Dirac bands}
\label{sec3}

The plasmon branches are obtained as the locations on the $(q,\omega)$-plane where the dielectric function $\epsilon(q,\omega \, \vert \, E_F)$ of a 
material becomes equal to zero. Within the random phase approximation, the dielectric function is calculated 

\begin{equation}
\label{epsmain}
\epsilon(\bar{\bf q}, \omega \, \vert \, E_F) = 1 - v_c(\bar{\bf q}) \, \Pi(\bar{\bf q}, \omega\, \vert \, E_F) \, , 
\end{equation}
where $v_c(q) = e^2 /(2 \epsilon_0 \epsilon_r \, q)$ is a Fourier-transformed Coulomb potential of the electron-electron interaction in a two-dimensional lattice,
$\epsilon_r$ is the relative dielectric constant of the SDB sheet which essentially depends on the dielectric substrate and 
$\Pi^{(0)}(q,\omega \, \vert \, E_F)$ is the polarization function.

\medskip 
Within the random phase approximation, the polarization function is calculated in the following way

\begin{eqnarray}
\label{Pi00}
&& \Pi^{(0)}(q_x, q_y, \omega\, \vert \, E_F) = \frac{1}{4 \pi^2} \, \sum \limits_{\xi = \pm 1} 
\int d k_x \, \int d k_y \, \sum \limits_{\lambda,\lambda' = \pm 1} \, \mbb{O}_{\lambda_1,\lambda_2} (\bar{\bf k}, \bar{\bf q} \,\vert \, 
a, \tau
) \, \times \\
\nonumber 
&& \times \left\{
\frac{
n_F [\varepsilon_{\lambda_1, \xi} (\bar{\bf k} \, \vert \, a, \tau) \, \vert \, \mu(T,E_F), T] - 
n_F [\varepsilon_{\lambda_2, \xi} (\bar{\bf k} + \bar{\bf q}  \, \vert \, a, \tau) \, \vert \, \mu(T,E_F), T]
}{
\hbar \omega + i 0^ + \varepsilon_{\lambda_1, \xi} (\bar{\bf k} \, \vert \, a, \tau) - 
\varepsilon_{\lambda_2, \xi} (\bar{\bf k} + \bar{\bf q}  \, \vert \, a, \tau)
}
\right\} \, , 
\end{eqnarray}
where $n_F [\varepsilon_{\lambda_1, \xi} (\bar{\bf k} \, \vert \, a, \tau) \, \vert \, \mu(T,E_F), T] = (1+ \tet{exp}[(\varepsilon_{\lambda_1, \xi} (\bar{\bf k} \, \vert \, a, \tau) - \mu)/(k_B T)])^{-1}$ is the Fermi-Dirac distribution function such that for a zero temperature it is reduced to a Heaviside step function
$n_F [\varepsilon_{\lambda_1, \xi} (\bar{\bf k} \, \vert \, a, \tau) \, \vert \, \mu(T,E_F), T] \longrightarrow \Theta[E_F - \varepsilon_{\lambda_1, \xi} (\bar{\bf k} \, \vert \, a, \tau) ]$.  The overlap factor $\mbb{O}_{\lambda_1,\lambda_2} (\bar{\bf k}, \bar{\bf q} \,\vert \, a, \tau)$ is defined as the wave function
overlap between the electron states in different bands and is calculated as 

\begin{equation}
\label{overlf}
\mbb{O}_{\lambda_1,\lambda_2} (\bar{\bf k}, \bar{\bf q} \,\vert \, a, \tau) = \Big \langle \Psi_{\lambda_1} \, ( \bar{\bf k} \, \vert \, a, \tau ) \Big \vert 
\Psi_{\lambda_2} \, (\bar{\bf k} + \bar{\bf q} \, \vert \, a, \tau ) \Big \rangle
\end{equation}

Using representation \eqref{wf02} of wave functions \eqref{wf01} corresponding to wave vectors $\bar{\bf k}$ and $\bar{\bf k} + \bar{\bf q}$, we immediately 
rewrite overlap factor \eqref{overlf} as 

\begin{eqnarray}
\label{overl021}
&& \mbb{O}_{\lambda_1,\lambda_2} (\bar{\bf k}, \bar{\bf q} \,\vert \, a, \tau) = \frac{1}{2} \left\{
1 + \lambda_1 \lambda_2  \frac{a \, k_x^2 + i v_F \, k_y}{(\hbar v_F \, k_y)^2 + (a \, k_x^2)^2} \, 
 \frac{a \, (k_x + q_x)^2 + i v_F \, (k_x + q_x)}{[\hbar v_F \, (k_y + q_y)]^2 + [a \, (k_y + q_y)^2]^2} 
\right\}
= \\
\nonumber 
&& = \frac{1}{2} \left\{ 1 + \lambda_1 \lambda_2 \cos \Theta_{
\left(
\mc{E}_{\bar{\bf k}}, \mc{E}_{\bar{\bf k} + \bar{\bf q}} 
\right)
}
\right\} =  \frac{1}{2} \left\{ 1 + \lambda_1 \lambda_2 
\frac{
\mc{E}_{\bar{\bf k}} + \mc{E}_{\bar{\bf q}} \,  \cos \Theta_{
\left(
\mc{E}_{\bar{\bf k}}, \mc{E}_{\bar{\bf q}} 
\right)
}
}{
\sqrt{\mc{E}_{\bar{\bf k}}^2 + \mc{E}_{\bar{\bf q}}^2 + 2 \mc{E}_{\bar{\bf k}} \mc{E}_{\bar{\bf q}} \cos \Theta_{
\left(
\mc{E}_{\bar{\bf k}}, \mc{E}_{\bar{\bf q}} 
\right)
}}
}
\right\}
\end{eqnarray}

Overlap factors $\mbb{O}_{\lambda_1,\lambda_2} (\bar{\bf k}, \bar{\bf q} \,\vert \, a, \tau) $ shown in Fig.~\ref{FIG:5} demonstrate a non-trivial dependence on both the magnitude and direction of wave vector shift $\bar{\bf q}$. which is different from that in graphene and most of the other known materials. However,  overlap $\mbb{O}_{\lambda_1,\lambda_2} (\bar{\bf k}, \bar{\bf q} \,\vert \, a, \tau) $ in Eq.~\eqref{overl021} could be presented in terms of a single angle $\Theta_{
\left(
\mc{E}_{\bar{\bf k}}, \mc{E}_{\bar{\bf k} + \bar{\bf q}} 
\right)
}$  and, therefore, the inter- ($\lambda_1 \lambda_2 = -1$) and intra-band ($\lambda_1 \lambda_2 = 1$) overlaps demonstrate completely opposite angular behavior.

\medskip
%

The real and imaginary parts of polarization function \eqref{Pi00} are presented in Figs.~\ref{FIG:6} and \ref{FIG:7}. The plasmon dispersions are obtained from equation \eqref{epsmain} as the zeros of dielectric function $\epsilon(\bar{\bf q}, \omega \, \vert \, E_F)$.  The real part of polarization function $ \Pi^{(0)}(q_x, q_y, \omega\, \vert \, E_F) $ plays a crucial role in shaping out the plasmon branches, while the imaginary part plays a crucial role in determining the plasmon damping and (inverse) lifetime since a plasmon could be only considered stable if $ \text{Im} \left[ \Pi^{(0)}(q_x, q_y, \omega\, \vert \, E_F) \right] \longrightarrow 0$  and $\vert \epsilon(\bar{\bf q}, \omega \, \vert \, E_F) \vert \longrightarrow 0$.

\par

We see that for a finite  transverse momentum component $q_y$  the results for both real and imaginary parts of polarization function  $ \Pi^{(0)}(q_x, q_y, \omega\, \vert \, E_F) $ are changed significantly,  but in both cases the real part of the polarization function could be found both  positive and negative which ensures that the plasmon actually exists.

\medskip 

Since the energy dispersions of semi-Dirac bands have no energy gap,  the region of an undamped plasmon is localized to the relatively small values of the wave vector $\bar{\bf q}$ and frequency $\omega$. At the same time, we clearly see a well-defined plasmon with zero or small $ \text{Im} \left[ \Pi^{(0)}(q_x, q_y, \omega\, \vert \, E_F) \right] $. A curved and nearly parabolic boundary of the particle-hole excitation region clearly resembles the plasmons in a two- dimensional electron gas (2DEG). This situation is expected because of parabolic dispersions in SDB’s in the $k_x$-direction.

\par

The plasmon branches presented in Fig.~\ref{FIG:8}  demonstrate a standard $\backsim \sqrt{q}$ behavior for $q_y = 0$, However, for a finite $q_y$  the branches are not monotonic and could be even decreasing with increasing $q_x$ with a clear minimum, which is the result of both tilting and anisotropy.

\medskip 

One of the most interesting features of the plasmons in semi-Dirac bands is their dependence on the tilting $\tau$. The plasmons in two-dimensional over-tilted Dirac bands with both linear dispersions and a gap were briefly analyzed in Ref.~[\onlinecite{yan2022anomalous}] in which additional branches were reported for critical in over-critical tilting. However, we don't know much about the damping and the lifetimes of these plasmons. It is crucial to study the tilting in connection with the parabolic dispersions in semi-Dirac bands and the corresponding unique schematics of the electronic transitions. Most importantly, the electron doping density for a given Fermi level is increased for an increasing tilting, as well as the area of the Fermi surface between the occupied and free electronic states. This situation is definitely expected to lead to an increased frequency (the energy) of the plasmon branches for a given wave vector $\bar{\bf q}$, which we definitely observe in Fig.~\ref{FIG:9}. It is also interesting to see that the polarization functions in the regions of low-damped plasmons (small $q$) demonstrate qualitatively the same behavior for different values of tilting parameter $\tau$.

\section{Summary and remarks}
\label{sec4}

In this paper, we have calculated the polarization function, plasmon dispersions and their damping for sem-Dirac bands. The energy band structure of this novel material is linear in the $k_y$-direction and parabolic along the $k_x$ axis, and has a zero energy band gap.

\medskip
The band structure of semi-Dirac bands also allows a finite $(0 < \tau < 1)$, critical $(\tau = 1)$  and over-critical tilting $(\tau > 1)$ so that one of the slopes in the $k_y$-direction could become zero or even negative. As a result, the area of the Fermo surface -- the contact surface between the free and occupied electron states at the Fermi level -- could increase and even become infinite for $\tau \rightarrow 1$  which substantially affects all the electronic and collective properties of SDB’s. In this case, a Fermi interface and a plasmon also exists even for zero Fermi level.

\medskip
We have obtained a well-defined, low-damped and anisotropic plasmon branch for both zero and a finite tilting in semi-Dirac bands. The boundary of the particle-hole modes or single particle excitation spectrum region -- locations in the $(q, \omega)$ plane in which a plasmon would decay into single-particle excitations --  is represented by a curved, nearly parabolic line, similarly to that in a two-dimensional electron gas (2DEG). A finite tilting $\tau > 0$ leads to the increase of the plasma frequency for a given wave vector $q$ and an extension of the region with low-damped plasmons. 

\par
We are confident that our finding and, specifically, demonstrating an existence of a low-damped plasmon in this new class of two-dimensional materials with non-trivial semi-Dirac dispersions and an earlier unseen schematics of the electronic transitions will undoubtedly find its numerous applications in creating of new nanoscale electronic devices, as well as general and theoretical condensed matter physics.

\begin{acknowledgements}
A.I. was supported by the funding received from TRADA-53-130, PSC-CUNY Award \# 65094-00 53. D.H. would like to acknowledge the Air Force Office of Scientific Research (AFOSR). G.G. was supported by Grant No. FA9453-21-1-0046 from the Air Force Research Laboratory (AFRL).
\end{acknowledgements}

\bibliography{PlaSDB}

\begin{thebibliography}{76}
\expandafter\ifx\csname natexlab\endcsname\relax\def\natexlab#1{#1}\fi
\expandafter\ifx\csname bibnamefont\endcsname\relax
  \def\bibnamefont#1{#1}\fi
\expandafter\ifx\csname bibfnamefont\endcsname\relax
  \def\bibfnamefont#1{#1}\fi
\expandafter\ifx\csname citenamefont\endcsname\relax
  \def\citenamefont#1{#1}\fi
\expandafter\ifx\csname url\endcsname\relax
  \def\url#1{\texttt{#1}}\fi
\expandafter\ifx\csname urlprefix\endcsname\relax\def\urlprefix{URL }\fi
\providecommand{\bibinfo}[2]{#2}
\providecommand{\eprint}[2][]{\url{#2}}

\bibitem[{\citenamefont{Tabert and Nicol}(2013)}]{tabert2013valley}
\bibinfo{author}{\bibfnamefont{C.~J.} \bibnamefont{Tabert}} \bibnamefont{and}
  \bibinfo{author}{\bibfnamefont{E.~J.} \bibnamefont{Nicol}},
  \bibinfo{journal}{Physical Review Letters} \textbf{\bibinfo{volume}{110}},
  \bibinfo{pages}{197402} (\bibinfo{year}{2013}).

\bibitem[{\citenamefont{Iurov et~al.}(2020{\natexlab{a}})\citenamefont{Iurov,
  Zhemchuzhna, Fekete, Gumbs, and Huang}}]{iurov2020klein}
\bibinfo{author}{\bibfnamefont{A.}~\bibnamefont{Iurov}},
  \bibinfo{author}{\bibfnamefont{L.}~\bibnamefont{Zhemchuzhna}},
  \bibinfo{author}{\bibfnamefont{P.}~\bibnamefont{Fekete}},
  \bibinfo{author}{\bibfnamefont{G.}~\bibnamefont{Gumbs}}, \bibnamefont{and}
  \bibinfo{author}{\bibfnamefont{D.}~\bibnamefont{Huang}},
  \bibinfo{journal}{Physical Review Research} \textbf{\bibinfo{volume}{2}},
  \bibinfo{pages}{043245} (\bibinfo{year}{2020}{\natexlab{a}}).

\bibitem[{\citenamefont{Islam and Basu}(2023)}]{islam2023properties}
\bibinfo{author}{\bibfnamefont{M.}~\bibnamefont{Islam}} \bibnamefont{and}
  \bibinfo{author}{\bibfnamefont{S.}~\bibnamefont{Basu}},
  \bibinfo{journal}{Journal of Physics: Condensed Matter}
  (\bibinfo{year}{2023}).

\bibitem[{\citenamefont{Iurov et~al.}(2019)\citenamefont{Iurov, Gumbs, and
  Huang}}]{iurov2019peculiar}
\bibinfo{author}{\bibfnamefont{A.}~\bibnamefont{Iurov}},
  \bibinfo{author}{\bibfnamefont{G.}~\bibnamefont{Gumbs}}, \bibnamefont{and}
  \bibinfo{author}{\bibfnamefont{D.}~\bibnamefont{Huang}},
  \bibinfo{journal}{Physical Review B} \textbf{\bibinfo{volume}{99}},
  \bibinfo{pages}{205135} (\bibinfo{year}{2019}).

\bibitem[{\citenamefont{Tamang and Biswas}(2023)}]{tamang2023probing}
\bibinfo{author}{\bibfnamefont{L.}~\bibnamefont{Tamang}} \bibnamefont{and}
  \bibinfo{author}{\bibfnamefont{T.}~\bibnamefont{Biswas}},
  \bibinfo{journal}{Physical Review B} \textbf{\bibinfo{volume}{107}},
  \bibinfo{pages}{085408} (\bibinfo{year}{2023}).

\bibitem[{\citenamefont{Islam et~al.}(2023{\natexlab{a}})\citenamefont{Islam,
  Biswas, and Basu}}]{islam2023effect}
\bibinfo{author}{\bibfnamefont{M.}~\bibnamefont{Islam}},
  \bibinfo{author}{\bibfnamefont{T.}~\bibnamefont{Biswas}}, \bibnamefont{and}
  \bibinfo{author}{\bibfnamefont{S.}~\bibnamefont{Basu}},
  \bibinfo{journal}{Physical Review B} \textbf{\bibinfo{volume}{108}},
  \bibinfo{pages}{085423} (\bibinfo{year}{2023}{\natexlab{a}}).

\bibitem[{\citenamefont{Gorbar et~al.}(2019)\citenamefont{Gorbar, Gusynin, and
  Oriekhov}}]{gorbar2019electron}
\bibinfo{author}{\bibfnamefont{E.}~\bibnamefont{Gorbar}},
  \bibinfo{author}{\bibfnamefont{V.}~\bibnamefont{Gusynin}}, \bibnamefont{and}
  \bibinfo{author}{\bibfnamefont{D.}~\bibnamefont{Oriekhov}},
  \bibinfo{journal}{Physical Review B} \textbf{\bibinfo{volume}{99}},
  \bibinfo{pages}{155124} (\bibinfo{year}{2019}).

\bibitem[{\citenamefont{Illes and Nicol}(2017)}]{illes2017klein}
\bibinfo{author}{\bibfnamefont{E.}~\bibnamefont{Illes}} \bibnamefont{and}
  \bibinfo{author}{\bibfnamefont{E.}~\bibnamefont{Nicol}},
  \bibinfo{journal}{Physical Review B} \textbf{\bibinfo{volume}{95}},
  \bibinfo{pages}{235432} (\bibinfo{year}{2017}).

\bibitem[{\citenamefont{Yan et~al.}(2023)\citenamefont{Yan, Tan, Guo, Chang
  et~al.}}]{yan2023highly}
\bibinfo{author}{\bibfnamefont{C.-X.} \bibnamefont{Yan}},
  \bibinfo{author}{\bibfnamefont{C.-Y.} \bibnamefont{Tan}},
  \bibinfo{author}{\bibfnamefont{H.}~\bibnamefont{Guo}},
  \bibinfo{author}{\bibfnamefont{H.-R.} \bibnamefont{Chang}},
  \bibnamefont{et~al.}, \bibinfo{journal}{Physical Review B}
  \textbf{\bibinfo{volume}{108}}, \bibinfo{pages}{195427}
  (\bibinfo{year}{2023}).

\bibitem[{\citenamefont{Iurov et~al.}(2020{\natexlab{b}})\citenamefont{Iurov,
  Zhemchuzhna, Dahal, Gumbs, and Huang}}]{iurov2020quantum}
\bibinfo{author}{\bibfnamefont{A.}~\bibnamefont{Iurov}},
  \bibinfo{author}{\bibfnamefont{L.}~\bibnamefont{Zhemchuzhna}},
  \bibinfo{author}{\bibfnamefont{D.}~\bibnamefont{Dahal}},
  \bibinfo{author}{\bibfnamefont{G.}~\bibnamefont{Gumbs}}, \bibnamefont{and}
  \bibinfo{author}{\bibfnamefont{D.}~\bibnamefont{Huang}},
  \bibinfo{journal}{Physical Review B} \textbf{\bibinfo{volume}{101}},
  \bibinfo{pages}{035129} (\bibinfo{year}{2020}{\natexlab{b}}).

\bibitem[{\citenamefont{Tan et~al.}(2021)\citenamefont{Tan, Yan, Zhao, Guo,
  Chang et~al.}}]{tan2021anisotropic}
\bibinfo{author}{\bibfnamefont{C.-Y.} \bibnamefont{Tan}},
  \bibinfo{author}{\bibfnamefont{C.-X.} \bibnamefont{Yan}},
  \bibinfo{author}{\bibfnamefont{Y.-H.} \bibnamefont{Zhao}},
  \bibinfo{author}{\bibfnamefont{H.}~\bibnamefont{Guo}},
  \bibinfo{author}{\bibfnamefont{H.-R.} \bibnamefont{Chang}},
  \bibnamefont{et~al.}, \bibinfo{journal}{Physical Review B}
  \textbf{\bibinfo{volume}{103}}, \bibinfo{pages}{125425}
  (\bibinfo{year}{2021}).

\bibitem[{\citenamefont{Gomes and Ramos}(2021)}]{gomes2021tilted}
\bibinfo{author}{\bibfnamefont{Y.}~\bibnamefont{Gomes}} \bibnamefont{and}
  \bibinfo{author}{\bibfnamefont{R.~O.} \bibnamefont{Ramos}},
  \bibinfo{journal}{Physical Review B} \textbf{\bibinfo{volume}{104}},
  \bibinfo{pages}{245111} (\bibinfo{year}{2021}).

\bibitem[{\citenamefont{Carbotte et~al.}(2019)\citenamefont{Carbotte, Bryenton,
  and Nicol}}]{carbotte2019optical}
\bibinfo{author}{\bibfnamefont{J.}~\bibnamefont{Carbotte}},
  \bibinfo{author}{\bibfnamefont{K.}~\bibnamefont{Bryenton}}, \bibnamefont{and}
  \bibinfo{author}{\bibfnamefont{E.}~\bibnamefont{Nicol}},
  \bibinfo{journal}{Physical Review B} \textbf{\bibinfo{volume}{99}},
  \bibinfo{pages}{115406} (\bibinfo{year}{2019}).

\bibitem[{\citenamefont{Islam and Saha}(2018)}]{islam2018driven}
\bibinfo{author}{\bibfnamefont{S.~F.} \bibnamefont{Islam}} \bibnamefont{and}
  \bibinfo{author}{\bibfnamefont{A.}~\bibnamefont{Saha}},
  \bibinfo{journal}{Physical Review B} \textbf{\bibinfo{volume}{98}},
  \bibinfo{pages}{235424} (\bibinfo{year}{2018}).

\bibitem[{\citenamefont{Xiong et~al.}(2023)\citenamefont{Xiong, Ba, Duan, Deng,
  Wang, and Wang}}]{xiong2023optical}
\bibinfo{author}{\bibfnamefont{Q.-Y.} \bibnamefont{Xiong}},
  \bibinfo{author}{\bibfnamefont{J.-Y.} \bibnamefont{Ba}},
  \bibinfo{author}{\bibfnamefont{H.-J.} \bibnamefont{Duan}},
  \bibinfo{author}{\bibfnamefont{M.-X.} \bibnamefont{Deng}},
  \bibinfo{author}{\bibfnamefont{Y.-M.} \bibnamefont{Wang}}, \bibnamefont{and}
  \bibinfo{author}{\bibfnamefont{R.-Q.} \bibnamefont{Wang}},
  \bibinfo{journal}{Physical Review B} \textbf{\bibinfo{volume}{107}},
  \bibinfo{pages}{155150} (\bibinfo{year}{2023}).

\bibitem[{\citenamefont{Mondal et~al.}(2022)\citenamefont{Mondal, Ganguly, and
  Basu}}]{mondal2022topology}
\bibinfo{author}{\bibfnamefont{S.}~\bibnamefont{Mondal}},
  \bibinfo{author}{\bibfnamefont{S.}~\bibnamefont{Ganguly}}, \bibnamefont{and}
  \bibinfo{author}{\bibfnamefont{S.}~\bibnamefont{Basu}},
  \bibinfo{journal}{Physical Sciences Reviews}  (\bibinfo{year}{2022}).

\bibitem[{\citenamefont{Shitrit et~al.}(2013)\citenamefont{Shitrit, Yulevich,
  Kleiner, and Hasman}}]{shitrit2013spin}
\bibinfo{author}{\bibfnamefont{N.}~\bibnamefont{Shitrit}},
  \bibinfo{author}{\bibfnamefont{I.}~\bibnamefont{Yulevich}},
  \bibinfo{author}{\bibfnamefont{V.}~\bibnamefont{Kleiner}}, \bibnamefont{and}
  \bibinfo{author}{\bibfnamefont{E.}~\bibnamefont{Hasman}},
  \bibinfo{journal}{Applied Physics Letters} \textbf{\bibinfo{volume}{103}}
  (\bibinfo{year}{2013}).

\bibitem[{\citenamefont{Shih et~al.}(2022)\citenamefont{Shih, Gumbs, Huang,
  Iurov, and Abranyos}}]{shih2022blocked}
\bibinfo{author}{\bibfnamefont{P.-H.} \bibnamefont{Shih}},
  \bibinfo{author}{\bibfnamefont{G.}~\bibnamefont{Gumbs}},
  \bibinfo{author}{\bibfnamefont{D.}~\bibnamefont{Huang}},
  \bibinfo{author}{\bibfnamefont{A.}~\bibnamefont{Iurov}}, \bibnamefont{and}
  \bibinfo{author}{\bibfnamefont{Y.}~\bibnamefont{Abranyos}},
  \bibinfo{journal}{Journal of Applied Physics} \textbf{\bibinfo{volume}{132}}
  (\bibinfo{year}{2022}).

\bibitem[{\citenamefont{Wang}(2005)}]{wang2005plasmon}
\bibinfo{author}{\bibfnamefont{X.-F.} \bibnamefont{Wang}},
  \bibinfo{journal}{Physical Review B} \textbf{\bibinfo{volume}{72}},
  \bibinfo{pages}{085317} (\bibinfo{year}{2005}).

\bibitem[{\citenamefont{Kristinsson et~al.}(2016)\citenamefont{Kristinsson,
  Kibis, Morina, and Shelykh}}]{kristinsson2016control}
\bibinfo{author}{\bibfnamefont{K.}~\bibnamefont{Kristinsson}},
  \bibinfo{author}{\bibfnamefont{O.~V.} \bibnamefont{Kibis}},
  \bibinfo{author}{\bibfnamefont{S.}~\bibnamefont{Morina}}, \bibnamefont{and}
  \bibinfo{author}{\bibfnamefont{I.~A.} \bibnamefont{Shelykh}},
  \bibinfo{journal}{Scientific reports} \textbf{\bibinfo{volume}{6}},
  \bibinfo{pages}{1} (\bibinfo{year}{2016}).

\bibitem[{\citenamefont{Kibis}(2010)}]{kibis2010metal}
\bibinfo{author}{\bibfnamefont{O.}~\bibnamefont{Kibis}},
  \bibinfo{journal}{Physical Review B} \textbf{\bibinfo{volume}{81}},
  \bibinfo{pages}{165433} (\bibinfo{year}{2010}).

\bibitem[{\citenamefont{Politano and Chiarello}(2014)}]{politano2014plasmon}
\bibinfo{author}{\bibfnamefont{A.}~\bibnamefont{Politano}} \bibnamefont{and}
  \bibinfo{author}{\bibfnamefont{G.}~\bibnamefont{Chiarello}},
  \bibinfo{journal}{Nanoscale} \textbf{\bibinfo{volume}{6}},
  \bibinfo{pages}{10927} (\bibinfo{year}{2014}).

\bibitem[{\citenamefont{Hwang and Sarma}(2007)}]{hwang2007dielectric}
\bibinfo{author}{\bibfnamefont{E.}~\bibnamefont{Hwang}} \bibnamefont{and}
  \bibinfo{author}{\bibfnamefont{S.~D.} \bibnamefont{Sarma}},
  \bibinfo{journal}{Physical Review B} \textbf{\bibinfo{volume}{75}},
  \bibinfo{pages}{205418} (\bibinfo{year}{2007}).

\bibitem[{\citenamefont{Polini et~al.}(2008)\citenamefont{Polini, Asgari,
  Borghi, Barlas, Pereg-Barnea, and MacDonald}}]{polini2008plasmons}
\bibinfo{author}{\bibfnamefont{M.}~\bibnamefont{Polini}},
  \bibinfo{author}{\bibfnamefont{R.}~\bibnamefont{Asgari}},
  \bibinfo{author}{\bibfnamefont{G.}~\bibnamefont{Borghi}},
  \bibinfo{author}{\bibfnamefont{Y.}~\bibnamefont{Barlas}},
  \bibinfo{author}{\bibfnamefont{T.}~\bibnamefont{Pereg-Barnea}},
  \bibnamefont{and}
  \bibinfo{author}{\bibfnamefont{A.}~\bibnamefont{MacDonald}},
  \bibinfo{journal}{Physical Review B} \textbf{\bibinfo{volume}{77}},
  \bibinfo{pages}{081411} (\bibinfo{year}{2008}).

\bibitem[{\citenamefont{Wunsch et~al.}(2006)\citenamefont{Wunsch, Stauber,
  Sols, and Guinea}}]{wunsch2006dynamical}
\bibinfo{author}{\bibfnamefont{B.}~\bibnamefont{Wunsch}},
  \bibinfo{author}{\bibfnamefont{T.}~\bibnamefont{Stauber}},
  \bibinfo{author}{\bibfnamefont{F.}~\bibnamefont{Sols}}, \bibnamefont{and}
  \bibinfo{author}{\bibfnamefont{F.}~\bibnamefont{Guinea}},
  \bibinfo{journal}{New Journal of Physics} \textbf{\bibinfo{volume}{8}},
  \bibinfo{pages}{318} (\bibinfo{year}{2006}).

\bibitem[{\citenamefont{Pyatkovskiy}(2008)}]{pyatkovskiy2008dynamical}
\bibinfo{author}{\bibfnamefont{P.}~\bibnamefont{Pyatkovskiy}},
  \bibinfo{journal}{Journal of Physics: Condensed Matter}
  \textbf{\bibinfo{volume}{21}}, \bibinfo{pages}{025506}
  (\bibinfo{year}{2008}).

\bibitem[{\citenamefont{Tabert and Nicol}(2014)}]{tabert2014dynamical}
\bibinfo{author}{\bibfnamefont{C.~J.} \bibnamefont{Tabert}} \bibnamefont{and}
  \bibinfo{author}{\bibfnamefont{E.~J.} \bibnamefont{Nicol}},
  \bibinfo{journal}{Physical Review B} \textbf{\bibinfo{volume}{89}},
  \bibinfo{pages}{195410} (\bibinfo{year}{2014}).

\bibitem[{\citenamefont{Iurov et~al.}(2017{\natexlab{a}})\citenamefont{Iurov,
  Huang, Gumbs, Pan, and Maradudin}}]{iurov2017effects}
\bibinfo{author}{\bibfnamefont{A.}~\bibnamefont{Iurov}},
  \bibinfo{author}{\bibfnamefont{D.}~\bibnamefont{Huang}},
  \bibinfo{author}{\bibfnamefont{G.}~\bibnamefont{Gumbs}},
  \bibinfo{author}{\bibfnamefont{W.}~\bibnamefont{Pan}}, \bibnamefont{and}
  \bibinfo{author}{\bibfnamefont{A.}~\bibnamefont{Maradudin}},
  \bibinfo{journal}{Physical Review B} \textbf{\bibinfo{volume}{96}},
  \bibinfo{pages}{081408} (\bibinfo{year}{2017}{\natexlab{a}}).

\bibitem[{\citenamefont{Yao et~al.}(2018)\citenamefont{Yao, Liu, Huang, Choi,
  Xie, Flor~Flores, Wu, Yu, Kwong, Huang et~al.}}]{yao2018broadband}
\bibinfo{author}{\bibfnamefont{B.}~\bibnamefont{Yao}},
  \bibinfo{author}{\bibfnamefont{Y.}~\bibnamefont{Liu}},
  \bibinfo{author}{\bibfnamefont{S.-W.} \bibnamefont{Huang}},
  \bibinfo{author}{\bibfnamefont{C.}~\bibnamefont{Choi}},
  \bibinfo{author}{\bibfnamefont{Z.}~\bibnamefont{Xie}},
  \bibinfo{author}{\bibfnamefont{J.}~\bibnamefont{Flor~Flores}},
  \bibinfo{author}{\bibfnamefont{Y.}~\bibnamefont{Wu}},
  \bibinfo{author}{\bibfnamefont{M.}~\bibnamefont{Yu}},
  \bibinfo{author}{\bibfnamefont{D.-L.} \bibnamefont{Kwong}},
  \bibinfo{author}{\bibfnamefont{Y.}~\bibnamefont{Huang}},
  \bibnamefont{et~al.}, \bibinfo{journal}{Nature Photonics}
  \textbf{\bibinfo{volume}{12}}, \bibinfo{pages}{22} (\bibinfo{year}{2018}).

\bibitem[{\citenamefont{Woessner et~al.}(2015)\citenamefont{Woessner,
  Lundeberg, Gao, Principi, Alonso-Gonz{\'a}lez, Carrega, Watanabe, Taniguchi,
  Vignale, Polini et~al.}}]{woessner2015highly}
\bibinfo{author}{\bibfnamefont{A.}~\bibnamefont{Woessner}},
  \bibinfo{author}{\bibfnamefont{M.~B.} \bibnamefont{Lundeberg}},
  \bibinfo{author}{\bibfnamefont{Y.}~\bibnamefont{Gao}},
  \bibinfo{author}{\bibfnamefont{A.}~\bibnamefont{Principi}},
  \bibinfo{author}{\bibfnamefont{P.}~\bibnamefont{Alonso-Gonz{\'a}lez}},
  \bibinfo{author}{\bibfnamefont{M.}~\bibnamefont{Carrega}},
  \bibinfo{author}{\bibfnamefont{K.}~\bibnamefont{Watanabe}},
  \bibinfo{author}{\bibfnamefont{T.}~\bibnamefont{Taniguchi}},
  \bibinfo{author}{\bibfnamefont{G.}~\bibnamefont{Vignale}},
  \bibinfo{author}{\bibfnamefont{M.}~\bibnamefont{Polini}},
  \bibnamefont{et~al.}, \bibinfo{journal}{Nature materials}
  \textbf{\bibinfo{volume}{14}}, \bibinfo{pages}{421} (\bibinfo{year}{2015}).

\bibitem[{\citenamefont{Li et~al.}(2017)\citenamefont{Li, Ren, and
  He}}]{li2017first}
\bibinfo{author}{\bibfnamefont{P.}~\bibnamefont{Li}},
  \bibinfo{author}{\bibfnamefont{X.}~\bibnamefont{Ren}}, \bibnamefont{and}
  \bibinfo{author}{\bibfnamefont{L.}~\bibnamefont{He}},
  \bibinfo{journal}{Physical Review B} \textbf{\bibinfo{volume}{96}},
  \bibinfo{pages}{165417} (\bibinfo{year}{2017}).

\bibitem[{\citenamefont{Iurov et~al.}(2017{\natexlab{b}})\citenamefont{Iurov,
  Gumbs, Huang, and Zhemchuzhna}}]{iurov2017controlling}
\bibinfo{author}{\bibfnamefont{A.}~\bibnamefont{Iurov}},
  \bibinfo{author}{\bibfnamefont{G.}~\bibnamefont{Gumbs}},
  \bibinfo{author}{\bibfnamefont{D.}~\bibnamefont{Huang}}, \bibnamefont{and}
  \bibinfo{author}{\bibfnamefont{L.}~\bibnamefont{Zhemchuzhna}},
  \bibinfo{journal}{Journal of Applied Physics} \textbf{\bibinfo{volume}{121}}
  (\bibinfo{year}{2017}{\natexlab{b}}).

\bibitem[{\citenamefont{Sarma and Li}(2013)}]{sarma2013intrinsic}
\bibinfo{author}{\bibfnamefont{S.~D.} \bibnamefont{Sarma}} \bibnamefont{and}
  \bibinfo{author}{\bibfnamefont{Q.}~\bibnamefont{Li}},
  \bibinfo{journal}{Physical Review B} \textbf{\bibinfo{volume}{87}},
  \bibinfo{pages}{235418} (\bibinfo{year}{2013}).

\bibitem[{\citenamefont{Iurov et~al.}(2017{\natexlab{c}})\citenamefont{Iurov,
  Gumbs, Huang, and Balakrishnan}}]{iurov2017thermal}
\bibinfo{author}{\bibfnamefont{A.}~\bibnamefont{Iurov}},
  \bibinfo{author}{\bibfnamefont{G.}~\bibnamefont{Gumbs}},
  \bibinfo{author}{\bibfnamefont{D.}~\bibnamefont{Huang}}, \bibnamefont{and}
  \bibinfo{author}{\bibfnamefont{G.}~\bibnamefont{Balakrishnan}},
  \bibinfo{journal}{Physical Review B} \textbf{\bibinfo{volume}{96}},
  \bibinfo{pages}{245403} (\bibinfo{year}{2017}{\natexlab{c}}).

\bibitem[{\citenamefont{Sarma and Madhukar}(1981)}]{sarma1981collective}
\bibinfo{author}{\bibfnamefont{S.~D.} \bibnamefont{Sarma}} \bibnamefont{and}
  \bibinfo{author}{\bibfnamefont{A.}~\bibnamefont{Madhukar}},
  \bibinfo{journal}{Physical Review B} \textbf{\bibinfo{volume}{23}},
  \bibinfo{pages}{805} (\bibinfo{year}{1981}).

\bibitem[{\citenamefont{Iurov et~al.}(2020{\natexlab{c}})\citenamefont{Iurov,
  Gumbs, and Huang}}]{iurov2020many}
\bibinfo{author}{\bibfnamefont{A.}~\bibnamefont{Iurov}},
  \bibinfo{author}{\bibfnamefont{G.}~\bibnamefont{Gumbs}}, \bibnamefont{and}
  \bibinfo{author}{\bibfnamefont{D.}~\bibnamefont{Huang}},
  \bibinfo{journal}{Journal of Physics: Condensed Matter}
  \textbf{\bibinfo{volume}{32}}, \bibinfo{pages}{415303}
  (\bibinfo{year}{2020}{\natexlab{c}}).

\bibitem[{\citenamefont{Henrard et~al.}(1999)\citenamefont{Henrard, Malengreau,
  Rudolf, Hevesi, Caudano, Lambin, and Cabioc’h}}]{henrard1999electron}
\bibinfo{author}{\bibfnamefont{L.}~\bibnamefont{Henrard}},
  \bibinfo{author}{\bibfnamefont{F.}~\bibnamefont{Malengreau}},
  \bibinfo{author}{\bibfnamefont{P.}~\bibnamefont{Rudolf}},
  \bibinfo{author}{\bibfnamefont{K.}~\bibnamefont{Hevesi}},
  \bibinfo{author}{\bibfnamefont{R.}~\bibnamefont{Caudano}},
  \bibinfo{author}{\bibfnamefont{P.}~\bibnamefont{Lambin}}, \bibnamefont{and}
  \bibinfo{author}{\bibfnamefont{T.}~\bibnamefont{Cabioc’h}},
  \bibinfo{journal}{Physical Review B} \textbf{\bibinfo{volume}{59}},
  \bibinfo{pages}{5832} (\bibinfo{year}{1999}).

\bibitem[{\citenamefont{Gumbs et~al.}(2014)\citenamefont{Gumbs, Balassis,
  Iurov, Fekete et~al.}}]{gumbs2014strongly}
\bibinfo{author}{\bibfnamefont{G.}~\bibnamefont{Gumbs}},
  \bibinfo{author}{\bibfnamefont{A.}~\bibnamefont{Balassis}},
  \bibinfo{author}{\bibfnamefont{A.}~\bibnamefont{Iurov}},
  \bibinfo{author}{\bibfnamefont{P.}~\bibnamefont{Fekete}},
  \bibnamefont{et~al.}, \bibinfo{journal}{The Scientific World Journal}
  \textbf{\bibinfo{volume}{2014}} (\bibinfo{year}{2014}).

\bibitem[{\citenamefont{Solov'Yov}(2005)}]{solov2005plasmon}
\bibinfo{author}{\bibfnamefont{A.~V.} \bibnamefont{Solov'Yov}},
  \bibinfo{journal}{International Journal of Modern Physics B}
  \textbf{\bibinfo{volume}{19}}, \bibinfo{pages}{4143} (\bibinfo{year}{2005}).

\bibitem[{\citenamefont{Ju et~al.}(1993)\citenamefont{Ju, Bulgac, and
  Keller}}]{ju1993excitation}
\bibinfo{author}{\bibfnamefont{N.}~\bibnamefont{Ju}},
  \bibinfo{author}{\bibfnamefont{A.}~\bibnamefont{Bulgac}}, \bibnamefont{and}
  \bibinfo{author}{\bibfnamefont{J.~W.} \bibnamefont{Keller}},
  \bibinfo{journal}{Physical Review B} \textbf{\bibinfo{volume}{48}},
  \bibinfo{pages}{9071} (\bibinfo{year}{1993}).

\bibitem[{\citenamefont{Brey and Fertig}(2007)}]{brey2007elementary}
\bibinfo{author}{\bibfnamefont{L.}~\bibnamefont{Brey}} \bibnamefont{and}
  \bibinfo{author}{\bibfnamefont{H.}~\bibnamefont{Fertig}},
  \bibinfo{journal}{Physical Review B} \textbf{\bibinfo{volume}{75}},
  \bibinfo{pages}{125434} (\bibinfo{year}{2007}).

\bibitem[{\citenamefont{Karimi and Knezevic}(2017)}]{karimi2017plasmons}
\bibinfo{author}{\bibfnamefont{F.}~\bibnamefont{Karimi}} \bibnamefont{and}
  \bibinfo{author}{\bibfnamefont{I.}~\bibnamefont{Knezevic}},
  \bibinfo{journal}{Physical Review B} \textbf{\bibinfo{volume}{96}},
  \bibinfo{pages}{125417} (\bibinfo{year}{2017}).

\bibitem[{\citenamefont{Gomez et~al.}(2016)\citenamefont{Gomez, Pisarra,
  Gravina, Pitarke, and Sindona}}]{gomez2016plasmon}
\bibinfo{author}{\bibfnamefont{C.~V.} \bibnamefont{Gomez}},
  \bibinfo{author}{\bibfnamefont{M.}~\bibnamefont{Pisarra}},
  \bibinfo{author}{\bibfnamefont{M.}~\bibnamefont{Gravina}},
  \bibinfo{author}{\bibfnamefont{J.~M.} \bibnamefont{Pitarke}},
  \bibnamefont{and} \bibinfo{author}{\bibfnamefont{A.}~\bibnamefont{Sindona}},
  \bibinfo{journal}{Physical review letters} \textbf{\bibinfo{volume}{117}},
  \bibinfo{pages}{116801} (\bibinfo{year}{2016}).

\bibitem[{\citenamefont{Iurov et~al.}(2021)\citenamefont{Iurov, Zhemchuzhna,
  Gumbs, Huang, Fekete, Anwar, Dahal, and Weekes}}]{iurov2021tailoring}
\bibinfo{author}{\bibfnamefont{A.}~\bibnamefont{Iurov}},
  \bibinfo{author}{\bibfnamefont{L.}~\bibnamefont{Zhemchuzhna}},
  \bibinfo{author}{\bibfnamefont{G.}~\bibnamefont{Gumbs}},
  \bibinfo{author}{\bibfnamefont{D.}~\bibnamefont{Huang}},
  \bibinfo{author}{\bibfnamefont{P.}~\bibnamefont{Fekete}},
  \bibinfo{author}{\bibfnamefont{F.}~\bibnamefont{Anwar}},
  \bibinfo{author}{\bibfnamefont{D.}~\bibnamefont{Dahal}}, \bibnamefont{and}
  \bibinfo{author}{\bibfnamefont{N.}~\bibnamefont{Weekes}},
  \bibinfo{journal}{Scientific reports} \textbf{\bibinfo{volume}{11}},
  \bibinfo{pages}{20577} (\bibinfo{year}{2021}).

\bibitem[{\citenamefont{Fei et~al.}(2015)\citenamefont{Fei, Goldflam, Wu, Dai,
  Wagner, McLeod, Liu, Post, Zhu, Janssen et~al.}}]{fei2015edge}
\bibinfo{author}{\bibfnamefont{Z.}~\bibnamefont{Fei}},
  \bibinfo{author}{\bibfnamefont{M.}~\bibnamefont{Goldflam}},
  \bibinfo{author}{\bibfnamefont{J.-S.} \bibnamefont{Wu}},
  \bibinfo{author}{\bibfnamefont{S.}~\bibnamefont{Dai}},
  \bibinfo{author}{\bibfnamefont{M.}~\bibnamefont{Wagner}},
  \bibinfo{author}{\bibfnamefont{A.}~\bibnamefont{McLeod}},
  \bibinfo{author}{\bibfnamefont{M.}~\bibnamefont{Liu}},
  \bibinfo{author}{\bibfnamefont{K.}~\bibnamefont{Post}},
  \bibinfo{author}{\bibfnamefont{S.}~\bibnamefont{Zhu}},
  \bibinfo{author}{\bibfnamefont{G.}~\bibnamefont{Janssen}},
  \bibnamefont{et~al.}, \bibinfo{journal}{Nano letters}
  \textbf{\bibinfo{volume}{15}}, \bibinfo{pages}{8271} (\bibinfo{year}{2015}).

\bibitem[{\citenamefont{Yan et~al.}(2012)\citenamefont{Yan, Li, Li, Zhu,
  Avouris, and Xia}}]{yan2012infrared}
\bibinfo{author}{\bibfnamefont{H.}~\bibnamefont{Yan}},
  \bibinfo{author}{\bibfnamefont{Z.}~\bibnamefont{Li}},
  \bibinfo{author}{\bibfnamefont{X.}~\bibnamefont{Li}},
  \bibinfo{author}{\bibfnamefont{W.}~\bibnamefont{Zhu}},
  \bibinfo{author}{\bibfnamefont{P.}~\bibnamefont{Avouris}}, \bibnamefont{and}
  \bibinfo{author}{\bibfnamefont{F.}~\bibnamefont{Xia}}, \bibinfo{journal}{Nano
  letters} \textbf{\bibinfo{volume}{12}}, \bibinfo{pages}{3766}
  (\bibinfo{year}{2012}).

\bibitem[{\citenamefont{Rold{\'a}n et~al.}(2009)\citenamefont{Rold{\'a}n,
  Fuchs, and Goerbig}}]{roldan2009collective}
\bibinfo{author}{\bibfnamefont{R.}~\bibnamefont{Rold{\'a}n}},
  \bibinfo{author}{\bibfnamefont{J.-N.} \bibnamefont{Fuchs}}, \bibnamefont{and}
  \bibinfo{author}{\bibfnamefont{M.}~\bibnamefont{Goerbig}},
  \bibinfo{journal}{Physical Review B} \textbf{\bibinfo{volume}{80}},
  \bibinfo{pages}{085408} (\bibinfo{year}{2009}).

\bibitem[{\citenamefont{Roldan et~al.}(2011)\citenamefont{Roldan, Goerbig, and
  Fuchs}}]{roldan2011theory}
\bibinfo{author}{\bibfnamefont{R.}~\bibnamefont{Roldan}},
  \bibinfo{author}{\bibfnamefont{M.}~\bibnamefont{Goerbig}}, \bibnamefont{and}
  \bibinfo{author}{\bibfnamefont{J.-N.} \bibnamefont{Fuchs}},
  \bibinfo{journal}{Physical Review B} \textbf{\bibinfo{volume}{83}},
  \bibinfo{pages}{205406} (\bibinfo{year}{2011}).

\bibitem[{\citenamefont{Balassis et~al.}(2020)\citenamefont{Balassis, Dahal,
  Gumbs, Iurov, Huang, and Roslyak}}]{balassis2020magnetoplasmons}
\bibinfo{author}{\bibfnamefont{A.}~\bibnamefont{Balassis}},
  \bibinfo{author}{\bibfnamefont{D.}~\bibnamefont{Dahal}},
  \bibinfo{author}{\bibfnamefont{G.}~\bibnamefont{Gumbs}},
  \bibinfo{author}{\bibfnamefont{A.}~\bibnamefont{Iurov}},
  \bibinfo{author}{\bibfnamefont{D.}~\bibnamefont{Huang}}, \bibnamefont{and}
  \bibinfo{author}{\bibfnamefont{O.}~\bibnamefont{Roslyak}},
  \bibinfo{journal}{Journal of Physics: Condensed Matter}
  \textbf{\bibinfo{volume}{32}}, \bibinfo{pages}{485301}
  (\bibinfo{year}{2020}).

\bibitem[{\citenamefont{Dutta et~al.}(2023)\citenamefont{Dutta, Chakraborty,
  and Agarwal}}]{dutta2023intrinsic}
\bibinfo{author}{\bibfnamefont{D.}~\bibnamefont{Dutta}},
  \bibinfo{author}{\bibfnamefont{A.}~\bibnamefont{Chakraborty}},
  \bibnamefont{and} \bibinfo{author}{\bibfnamefont{A.}~\bibnamefont{Agarwal}},
  \bibinfo{journal}{Physical Review B} \textbf{\bibinfo{volume}{107}},
  \bibinfo{pages}{165404} (\bibinfo{year}{2023}).

\bibitem[{\citenamefont{Tamang et~al.}(2023)\citenamefont{Tamang, Verma, and
  Biswas}}]{tamang2023orbital}
\bibinfo{author}{\bibfnamefont{L.}~\bibnamefont{Tamang}},
  \bibinfo{author}{\bibfnamefont{S.}~\bibnamefont{Verma}}, \bibnamefont{and}
  \bibinfo{author}{\bibfnamefont{T.}~\bibnamefont{Biswas}},
  \bibinfo{journal}{arXiv preprint arXiv:2309.07074}  (\bibinfo{year}{2023}).

\bibitem[{\citenamefont{Nimyi et~al.}(2022)\citenamefont{Nimyi, K{\"o}nye,
  Sharapov, and Gusynin}}]{nimyi2022landau}
\bibinfo{author}{\bibfnamefont{I.}~\bibnamefont{Nimyi}},
  \bibinfo{author}{\bibfnamefont{V.}~\bibnamefont{K{\"o}nye}},
  \bibinfo{author}{\bibfnamefont{S.}~\bibnamefont{Sharapov}}, \bibnamefont{and}
  \bibinfo{author}{\bibfnamefont{V.}~\bibnamefont{Gusynin}},
  \bibinfo{journal}{Physical Review B} \textbf{\bibinfo{volume}{106}},
  \bibinfo{pages}{085401} (\bibinfo{year}{2022}).

\bibitem[{\citenamefont{Oriekhov and Voronov}(2023)}]{oriekhov2023size}
\bibinfo{author}{\bibfnamefont{D.}~\bibnamefont{Oriekhov}} \bibnamefont{and}
  \bibinfo{author}{\bibfnamefont{S.}~\bibnamefont{Voronov}},
  \bibinfo{journal}{Journal of Physics: Condensed Matter}
  (\bibinfo{year}{2023}).

\bibitem[{\citenamefont{Farhat et~al.}(2013)\citenamefont{Farhat, Guenneau, and
  Ba{\u{g}}c{\i}}}]{farhat2013exciting}
\bibinfo{author}{\bibfnamefont{M.}~\bibnamefont{Farhat}},
  \bibinfo{author}{\bibfnamefont{S.}~\bibnamefont{Guenneau}}, \bibnamefont{and}
  \bibinfo{author}{\bibfnamefont{H.}~\bibnamefont{Ba{\u{g}}c{\i}}},
  \bibinfo{journal}{Physical review letters} \textbf{\bibinfo{volume}{111}},
  \bibinfo{pages}{237404} (\bibinfo{year}{2013}).

\bibitem[{\citenamefont{Brongersma et~al.}(2015)\citenamefont{Brongersma,
  Halas, and Nordlander}}]{brongersma2015plasmon}
\bibinfo{author}{\bibfnamefont{M.~L.} \bibnamefont{Brongersma}},
  \bibinfo{author}{\bibfnamefont{N.~J.} \bibnamefont{Halas}}, \bibnamefont{and}
  \bibinfo{author}{\bibfnamefont{P.}~\bibnamefont{Nordlander}},
  \bibinfo{journal}{Nature nanotechnology} \textbf{\bibinfo{volume}{10}},
  \bibinfo{pages}{25} (\bibinfo{year}{2015}).

\bibitem[{\citenamefont{Vinogradov et~al.}(2018)\citenamefont{Vinogradov,
  Dorofeenko, Pukhov, and Lisyansky}}]{vinogradov2018exciting}
\bibinfo{author}{\bibfnamefont{A.}~\bibnamefont{Vinogradov}},
  \bibinfo{author}{\bibfnamefont{A.}~\bibnamefont{Dorofeenko}},
  \bibinfo{author}{\bibfnamefont{A.}~\bibnamefont{Pukhov}}, \bibnamefont{and}
  \bibinfo{author}{\bibfnamefont{A.}~\bibnamefont{Lisyansky}},
  \bibinfo{journal}{Physical Review B} \textbf{\bibinfo{volume}{97}},
  \bibinfo{pages}{235407} (\bibinfo{year}{2018}).

\bibitem[{\citenamefont{Simon et~al.}(1983)\citenamefont{Simon, Short,
  Williams, and Dewandre}}]{simon1983inhomogeneous}
\bibinfo{author}{\bibfnamefont{A.}~\bibnamefont{Simon}},
  \bibinfo{author}{\bibfnamefont{R.}~\bibnamefont{Short}},
  \bibinfo{author}{\bibfnamefont{E.}~\bibnamefont{Williams}}, \bibnamefont{and}
  \bibinfo{author}{\bibfnamefont{T.}~\bibnamefont{Dewandre}},
  \bibinfo{journal}{The Physics of fluids} \textbf{\bibinfo{volume}{26}},
  \bibinfo{pages}{3107} (\bibinfo{year}{1983}).

\bibitem[{\citenamefont{Gumbs et~al.}(2015)\citenamefont{Gumbs, Iurov, Huang,
  and Pan}}]{gumbs2015tunable}
\bibinfo{author}{\bibfnamefont{G.}~\bibnamefont{Gumbs}},
  \bibinfo{author}{\bibfnamefont{A.}~\bibnamefont{Iurov}},
  \bibinfo{author}{\bibfnamefont{D.}~\bibnamefont{Huang}}, \bibnamefont{and}
  \bibinfo{author}{\bibfnamefont{W.}~\bibnamefont{Pan}},
  \bibinfo{journal}{Journal of Applied Physics} \textbf{\bibinfo{volume}{118}}
  (\bibinfo{year}{2015}).

\bibitem[{\citenamefont{Petrov et~al.}(2017)\citenamefont{Petrov, Svintsov,
  Ryzhii, and Shur}}]{petrov2017amplified}
\bibinfo{author}{\bibfnamefont{A.~S.} \bibnamefont{Petrov}},
  \bibinfo{author}{\bibfnamefont{D.}~\bibnamefont{Svintsov}},
  \bibinfo{author}{\bibfnamefont{V.}~\bibnamefont{Ryzhii}}, \bibnamefont{and}
  \bibinfo{author}{\bibfnamefont{M.~S.} \bibnamefont{Shur}},
  \bibinfo{journal}{Physical Review B} \textbf{\bibinfo{volume}{95}},
  \bibinfo{pages}{045405} (\bibinfo{year}{2017}).

\bibitem[{\citenamefont{Koseki et~al.}(2016)\citenamefont{Koseki, Ryzhii,
  Otsuji, Popov, and Satou}}]{koseki2016giant}
\bibinfo{author}{\bibfnamefont{Y.}~\bibnamefont{Koseki}},
  \bibinfo{author}{\bibfnamefont{V.}~\bibnamefont{Ryzhii}},
  \bibinfo{author}{\bibfnamefont{T.}~\bibnamefont{Otsuji}},
  \bibinfo{author}{\bibfnamefont{V.}~\bibnamefont{Popov}}, \bibnamefont{and}
  \bibinfo{author}{\bibfnamefont{A.}~\bibnamefont{Satou}},
  \bibinfo{journal}{Physical Review B} \textbf{\bibinfo{volume}{93}},
  \bibinfo{pages}{245408} (\bibinfo{year}{2016}).

\bibitem[{\citenamefont{Malcolm and Nicol}(2016)}]{malcolm2016frequency}
\bibinfo{author}{\bibfnamefont{J.}~\bibnamefont{Malcolm}} \bibnamefont{and}
  \bibinfo{author}{\bibfnamefont{E.}~\bibnamefont{Nicol}},
  \bibinfo{journal}{Physical Review B} \textbf{\bibinfo{volume}{93}},
  \bibinfo{pages}{165433} (\bibinfo{year}{2016}).

\bibitem[{\citenamefont{Oriekho}(2023)}]{oriekho2023quantum}
\bibinfo{author}{\bibfnamefont{D.}~\bibnamefont{Oriekho}}, Ph.D. thesis,
  \bibinfo{school}{Leiden University} (\bibinfo{year}{2023}).

\bibitem[{\citenamefont{Oriekhov and Gusynin}(2020)}]{oriekhov2020rkky}
\bibinfo{author}{\bibfnamefont{D.}~\bibnamefont{Oriekhov}} \bibnamefont{and}
  \bibinfo{author}{\bibfnamefont{V.}~\bibnamefont{Gusynin}},
  \bibinfo{journal}{Physical Review B} \textbf{\bibinfo{volume}{101}},
  \bibinfo{pages}{235162} (\bibinfo{year}{2020}).

\bibitem[{\citenamefont{Islam et~al.}(2023{\natexlab{b}})\citenamefont{Islam,
  Biswas, and Basu}}]{islam2023role}
\bibinfo{author}{\bibfnamefont{M.}~\bibnamefont{Islam}},
  \bibinfo{author}{\bibfnamefont{T.}~\bibnamefont{Biswas}}, \bibnamefont{and}
  \bibinfo{author}{\bibfnamefont{S.}~\bibnamefont{Basu}},
  \bibinfo{journal}{arXiv preprint arXiv:2304.08830}
  (\bibinfo{year}{2023}{\natexlab{b}}).

\bibitem[{\citenamefont{Stauber et~al.}(2013)\citenamefont{Stauber, San-Jose,
  and Brey}}]{stauber2013optical}
\bibinfo{author}{\bibfnamefont{T.}~\bibnamefont{Stauber}},
  \bibinfo{author}{\bibfnamefont{P.}~\bibnamefont{San-Jose}}, \bibnamefont{and}
  \bibinfo{author}{\bibfnamefont{L.}~\bibnamefont{Brey}}, \bibinfo{journal}{New
  Journal of Physics} \textbf{\bibinfo{volume}{15}}, \bibinfo{pages}{113050}
  (\bibinfo{year}{2013}).

\bibitem[{\citenamefont{Hayn et~al.}(2021)\citenamefont{Hayn, Wei, Silkin, and
  van~den Brink}}]{hayn2021plasmons}
\bibinfo{author}{\bibfnamefont{R.}~\bibnamefont{Hayn}},
  \bibinfo{author}{\bibfnamefont{T.}~\bibnamefont{Wei}},
  \bibinfo{author}{\bibfnamefont{V.~M.} \bibnamefont{Silkin}},
  \bibnamefont{and} \bibinfo{author}{\bibfnamefont{J.}~\bibnamefont{van~den
  Brink}}, \bibinfo{journal}{Physical Review Materials}
  \textbf{\bibinfo{volume}{5}}, \bibinfo{pages}{024201} (\bibinfo{year}{2021}).

\bibitem[{\citenamefont{Yan et~al.}(2022)\citenamefont{Yan, Zhang, Tan, Chang,
  Zhou, Yao, and Guo}}]{yan2022anomalous}
\bibinfo{author}{\bibfnamefont{C.-X.} \bibnamefont{Yan}},
  \bibinfo{author}{\bibfnamefont{F.}~\bibnamefont{Zhang}},
  \bibinfo{author}{\bibfnamefont{C.-Y.} \bibnamefont{Tan}},
  \bibinfo{author}{\bibfnamefont{H.-R.} \bibnamefont{Chang}},
  \bibinfo{author}{\bibfnamefont{J.}~\bibnamefont{Zhou}},
  \bibinfo{author}{\bibfnamefont{Y.}~\bibnamefont{Yao}}, \bibnamefont{and}
  \bibinfo{author}{\bibfnamefont{H.}~\bibnamefont{Guo}},
  \bibinfo{journal}{arXiv preprint arXiv:2211.11266}  (\bibinfo{year}{2022}).

\bibitem[{\citenamefont{Kajita et~al.}(2014)\citenamefont{Kajita, Nishio,
  Tajima, Suzumura, and Kobayashi}}]{kajita2014molecular}
\bibinfo{author}{\bibfnamefont{K.}~\bibnamefont{Kajita}},
  \bibinfo{author}{\bibfnamefont{Y.}~\bibnamefont{Nishio}},
  \bibinfo{author}{\bibfnamefont{N.}~\bibnamefont{Tajima}},
  \bibinfo{author}{\bibfnamefont{Y.}~\bibnamefont{Suzumura}}, \bibnamefont{and}
  \bibinfo{author}{\bibfnamefont{A.}~\bibnamefont{Kobayashi}},
  \bibinfo{journal}{Journal of the Physical Society of Japan}
  \textbf{\bibinfo{volume}{83}}, \bibinfo{pages}{072002}
  (\bibinfo{year}{2014}).

\bibitem[{\citenamefont{Sadhukhan and
  Agarwal}(2017)}]{sadhukhan2017anisotropic}
\bibinfo{author}{\bibfnamefont{K.}~\bibnamefont{Sadhukhan}} \bibnamefont{and}
  \bibinfo{author}{\bibfnamefont{A.}~\bibnamefont{Agarwal}},
  \bibinfo{journal}{Physical Review B} \textbf{\bibinfo{volume}{96}},
  \bibinfo{pages}{035410} (\bibinfo{year}{2017}).

\bibitem[{\citenamefont{Balassis et~al.}(2022)\citenamefont{Balassis, Gumbs,
  and Roslyak}}]{balassis2022polarizability}
\bibinfo{author}{\bibfnamefont{A.}~\bibnamefont{Balassis}},
  \bibinfo{author}{\bibfnamefont{G.}~\bibnamefont{Gumbs}}, \bibnamefont{and}
  \bibinfo{author}{\bibfnamefont{O.}~\bibnamefont{Roslyak}},
  \bibinfo{journal}{Physics Letters A} \textbf{\bibinfo{volume}{449}},
  \bibinfo{pages}{128353} (\bibinfo{year}{2022}).

\bibitem[{\citenamefont{Mojarro et~al.}(2022)\citenamefont{Mojarro,
  Carrillo-Bastos, and Maytorena}}]{mojarro2022hyperbolic}
\bibinfo{author}{\bibfnamefont{M.}~\bibnamefont{Mojarro}},
  \bibinfo{author}{\bibfnamefont{R.}~\bibnamefont{Carrillo-Bastos}},
  \bibnamefont{and} \bibinfo{author}{\bibfnamefont{J.~A.}
  \bibnamefont{Maytorena}}, \bibinfo{journal}{Physical Review B}
  \textbf{\bibinfo{volume}{105}}, \bibinfo{pages}{L201408}
  (\bibinfo{year}{2022}).

\bibitem[{\citenamefont{Mojarro et~al.}(2021)\citenamefont{Mojarro,
  Carrillo-Bastos, and Maytorena}}]{mojarro2021optical}
\bibinfo{author}{\bibfnamefont{M.}~\bibnamefont{Mojarro}},
  \bibinfo{author}{\bibfnamefont{R.}~\bibnamefont{Carrillo-Bastos}},
  \bibnamefont{and} \bibinfo{author}{\bibfnamefont{J.~A.}
  \bibnamefont{Maytorena}}, \bibinfo{journal}{Physical Review B}
  \textbf{\bibinfo{volume}{103}}, \bibinfo{pages}{165415}
  (\bibinfo{year}{2021}).

\bibitem[{\citenamefont{Jalali-Mola and Jafari}(2018)}]{jalali2018tilt}
\bibinfo{author}{\bibfnamefont{Z.}~\bibnamefont{Jalali-Mola}} \bibnamefont{and}
  \bibinfo{author}{\bibfnamefont{S.}~\bibnamefont{Jafari}},
  \bibinfo{journal}{Physical Review B} \textbf{\bibinfo{volume}{98}},
  \bibinfo{pages}{195415} (\bibinfo{year}{2018}).

\bibitem[{\citenamefont{Torbatian et~al.}(2021)\citenamefont{Torbatian, Novko,
  and Asgari}}]{torbatian2021hyperbolic}
\bibinfo{author}{\bibfnamefont{Z.}~\bibnamefont{Torbatian}},
  \bibinfo{author}{\bibfnamefont{D.}~\bibnamefont{Novko}}, \bibnamefont{and}
  \bibinfo{author}{\bibfnamefont{R.}~\bibnamefont{Asgari}},
  \bibinfo{journal}{Physical Review B} \textbf{\bibinfo{volume}{104}},
  \bibinfo{pages}{075432} (\bibinfo{year}{2021}).

\bibitem[{\citenamefont{Sadhukhan et~al.}(2020)\citenamefont{Sadhukhan,
  Politano, and Agarwal}}]{sadhukhan2020novel}
\bibinfo{author}{\bibfnamefont{K.}~\bibnamefont{Sadhukhan}},
  \bibinfo{author}{\bibfnamefont{A.}~\bibnamefont{Politano}}, \bibnamefont{and}
  \bibinfo{author}{\bibfnamefont{A.}~\bibnamefont{Agarwal}},
  \bibinfo{journal}{Physical Review Letters} \textbf{\bibinfo{volume}{124}},
  \bibinfo{pages}{046803} (\bibinfo{year}{2020}).

\bibitem[{\citenamefont{Dey and Ghosh}(2022)}]{dey2022dynamical}
\bibinfo{author}{\bibfnamefont{B.}~\bibnamefont{Dey}} \bibnamefont{and}
  \bibinfo{author}{\bibfnamefont{T.~K.} \bibnamefont{Ghosh}},
  \bibinfo{journal}{Journal of Physics: Condensed Matter}
  \textbf{\bibinfo{volume}{34}}, \bibinfo{pages}{255701}
  (\bibinfo{year}{2022}).

\end{thebibliography}
\end{document}